\documentclass[aps,showpacs,groupedaddress,eqsecnum,notitlepage,nofootinbib]{revtex4-2}
\usepackage{graphicx}
\usepackage{amsmath}
\usepackage{amsfonts}
\usepackage{amssymb}
\usepackage{enumerate, color}
\usepackage{subfigure}
\usepackage{color}
\usepackage[utf8]{inputenc}
\usepackage{float}
\usepackage[colorlinks=true,pdfstartview=FitV,linkcolor=blue,citecolor=blue,urlcolor=blue,breaklinks=true]{hyperref}
\usepackage{orcidlink}
\newcommand{\be}{\begin{equation}}
\newcommand{\ee}{\end{equation}}
\newcommand{\ben}{\begin{eqnarray}}
\newcommand{\een}{\end{eqnarray}}
\newcommand{\bes}{\begin{subequations}}
\newcommand{\ees}{\end{subequations}}

\begin{document}
\title{Charged Lifshitz black holes from general covariance breaking}
\author{D. C. Moreira$^{1}$\,\orcidlink{0000-0002-8799-3206}}
\author{A. S. Lemos$^{2}$\,\orcidlink{0000-0002-3940-0779}}
\author{F A. Brito$^{2,3}$\,\orcidlink{0000-0001-9465-6868}}
\affiliation{$^1$Centro de Ciências, Tecnologia e Saúde, Universidade Estadual da Paraíba, 58233-000, Araruna, PB, Brazil. }
\affiliation{$^2$Unidade Acadêmica de Física, Universidade Federal da Campina Grande, 58109-970, Campina Grande, PB, Brazil}
\affiliation{$^3$Departamento de F\'isica, Universidade Federal da Para\'iba, 58051-970, Jo\~ao Pessoa, PB, Brazil}

\email{moreira.dancesar@gmail.com\\adiellemos@gmail.com\\fabrito@df.ufcg.edu.br}
\begin{abstract}
In this work we use a general covariance breaking mechanism to obtain a class of charged black holes whose background geometry asymptotically approaches Lifshitz spacetimes. We discuss how this method affects Einstein's equations and explore the thermodynamics and critical behavior of the solution found.
\end{abstract}

\maketitle
\section{Introduction}
The Anti-de Sitter/Conformal Field Theory (AdS/CFT) correspondence introduced in theoretical physics a way to study strongly coupled relativistic field theories from the dynamics of gravitational models in a higher dimension \cite{maldacena1999large,aharony2000large}. Although it was originally arranged to perform String Theory calculations on asymptotically anti-de Sitter ($\text{AdS}_{D}$) spacetimes in the supergravity regime, nowadays it is understood that the dualities between (nongravitational) field theories and gravitational models - now generically known as Gauge/Gravity duality - can be extended to much broader sectors of physics \cite{ammon2015gauge,hartnoll2009lectures}. One of these sectors has been working on the search for applications of gauge/gravity duality in strongly coupled nonrelativistic systems in condensed matter physics. In particular, it was proposed in \cite{kachru2008gravity} that background geometries equipped with nonrelativistic coordinate scaling are useful in applying gauge/gravity duality to study quantum critical points in Lifshitz field theory. An example of such backgrounds is given by 
\begin{equation}\label{lifmetric}
ds^2=-\left(\frac{r}{\ell}\right)^{2z} dt^2+\left(\frac{\ell}{r}\right)^{2}dr^2+\left(\frac{r}{\ell}\right)^{2} dx^i dx^i,
\end{equation}
which has been explored in several applications involving nonrelativistic field theories \cite{taylor2016lifshitz,hartnoll2009lectures,zaanen2015holographic}. The metric \eqref{lifmetric} is invariant under the scaling \cite{taylor2016lifshitz}
\begin{equation}\label{lifscal}
\mathcal{D}_{z}: t\rightarrow \beta^z t, ~x^i\rightarrow \beta x^i~~\text{and} ~~ r\rightarrow r/\beta.
\end{equation}
where $z$ represents a {\it dynamical exponent}, which works as a measure of the spacetime anisotropy. Models that present the scaling invariance \eqref{lifscal} have gained prominence in field theory since for $z=2$ Lifshitz field theories are dual to the background metric \eqref{lifmetric}, enjoying the same set of nonrelativistic symmetries \cite{kachru2008gravity}, and for $z=3$ it is possible to find consistent proposals of well-behaved (renormalizable) gravity in the UV regime of Hor\"ava-Lifshitz (HL) models \cite{hovrava2009quantum,sotiriou2011hovrava}. Moreover, for $z=1$ we retrieve the relativistic setup and the $\text{AdS}_{D}$ geometry, but for $z\neq1$ the metric \eqref{lifmetric} usually emerges in the presence of massive vector fields \cite{taylor2008non} or within HL scenarios \cite{griffin2013lifshitz}.

A natural extension to the study of Lifshitz geometries is the search for black hole solutions asymptotically approaching the background \eqref{lifmetric}. Such solutions are used in systems where the temperature of the gravity side acts to involve the dual field theory in thermal bath \cite{taylor2008non,taylor2016lifshitz,balasubramanian2009analytic} and  due to this interest, neutral \cite{danielsson2009black,mann2009lifshitz,deveciouglu2014lifshitz,bertoldi2009black} and charged \cite{pang2010charged,brynjolfsson2013holographic,dehghani2011charged,alvarez2014nonlinearly} Lifshitz black holes were found and several other studies were carried out on this subject (see, for instance \cite{ayon2010analytic,ayon2009lifshitz,brito2020black,natsuume2018holographic,bazeia2015two,deveciouglu2011thermodynamics,gonzalez2011field,melnikov2019lifshitz,bravo2020thermodynamics,ayon2019microscopic,zangeneh2015thermodynamics,alvarez2014nonlinearly,liu2014thermodynamics,tarrio2011black,naeimipour2021lifshitz,tallarita2014holographic,li2014non,lu2014lifshitz,bravo2022lifshitz}). In particular, by extending the connection between the thermodynamics of charged $AdS$ black holes and the thermodynamics of liquid-gas systems \cite{chamblin1999charged,chamblin1999holography,dolan2011pressure,kubizvnak2012p} to charged Lifshitz black holes made it possible, for example, to use holographic methods in the analysis of critical phenomena in superconductors \cite{brynjolfsson2010holographic}, complexity \cite{zhu2020holographic}, electrical conductivity \cite{jain2010universal} and fermion systems \cite{brynjolfsson2010black,fang2012holographic} in models with no underlying conformal symmetry. 
 
Recently, as part of discussions on extending Derrick's theorem to curved backgrounds \cite{hobart1963instability,derrick1964comments,radmore1978non,palmer1979derrick,bazeia2003new,alestas2019evading,carloni2019derrick,morris2021radially,mandal2021solitons,morris2022bps,moreira2022analytical,moreira2022erratum,moreira2022scalar,moreira2023localized}, a first-order method for capturing analytical solutions of classical scalar fields in the probe regime was developed for Lifshitz spacetimes \cite{moreira2022analytical,moreira2022erratum} and extended to asymptotically Lifshitz backgrounds \cite{moreira2022scalar} as well as for a general class of radially symmetric, static geometries, including black holes \cite{moreira2023localized}. These models generalize the proposal presented in \cite{bazeia2003new} to circumvent Derricks' theorem on flat spacetimes, which is based on the requirement that the action be explicitly dependent on the background coordinates, resulting in an explicit diffeomorphism invariance breaking. As a whole, these results point out that if general covariance is set aside by imposing an explicitly coordinate-dependent scalar potential, one can find radially stable scalar fields on a large class of static, nonbackreacting backgrounds. Furthermore, the mechanism devised for finding analytical probe scalars is particularly useful here because it hints at how to conveniently write the scalar potential to analytically solve the full set of field equations in the presence of backreaction. 

Diffeomorphism invariance breaking mechanisms has been used in several scenarios, either  to build effective models or to explore deviations from the principle of general covariance. For example, the idea that gauge theories and General Relativity can be emergent, low-energy, effects of a more fundamental theory which does not respect diffeomorphism invariance was explored in \cite{anber2010breaking}, where traces of this fundamental theory could be found in fluctuations around General Relativity equipped with noncovariant terms. The spatial diffeomorphism breaking as a mechanism to study primordial power
spectrum from tensor fluctuations in inflationary models is discussed in \cite{cannone2015generalised,graef2015breaking,graef2017constraining}. In \cite{milgrom2019noncovariance}, general covariance violation is used to study aspects of galaxy behavior in the low-acceleration regime in modified Newtonian dynamics (MOND) models, whose emergent dynamics often exhibit significant discrepancies from Newtonian dynamics. Recently \cite{reyes2021hamiltonian, reyes2022modified}, studies involving nondynamical background fields which explicitly violate diffeomorphism invariance in modified gravity models have led to cosmological scenarios where the accelerated expansion of the Universe can be driven without the need for dark matter or energy \cite{reyes2022cosmology}. Other aspects of explicitly breaking diffeomorphism invariance in effective models by using nondynamical background fields are addressed in \cite{kostelecky2021backgrounds,bluhm2017gravity,bluhm2015spacetime,bluhm2019gravity,bluhm2021gravity,hidaka2015effective}.

A detailed study on spontaneous and explicit diffeomorphism breaking in gravitational models is presented in \cite{bluhm2015explicit}. In particular, when considering metric backreactions in the presence of fixed background fields, difficulties arise regarding the compatibility of Einstein's equations. The explicit coordinate dependence on the action denotes the presence of nondynamical degrees of freedom in the system, which breaks diffeomorphism invariance and generates a nonconserved energy-momentum tensor. It brings forth a problem in the field equations as the Einstein tensor is always conserved due to the contracted Bianchi identity. One can overcome this problem by setting the divergence of the energy-momentum tensor to zero by hand, which must hold on shell and allows solving the Einstein´s equations to find the appropriate background geometry, but introduces a constraint which eliminates degrees of freedom from the field solutions. This idea was explored in \cite{moreira2022scalar} for an Einstein-Maxwell-scalar system equipped with a nondynamical dielectric function and a scalar potential explicitly dependent on the background coordinates. In this case, the electric charge associated with the Maxwell field is eliminated from the metric solution due to the constraint imposed on the energy-momentum tensor to ensure its compatibility with Einstein's equations, generating a neutral Lifshitz black hole.

In this work we present the charged extension of the Lifshitz black hole solution presented in \cite{moreira2022scalar}. To this end, we add a second Maxwell field to the system whose function is to provide electrical charge to the background geometry. The two Maxwell fields present in the model differ from each other by the fact that one of them does not interact with the dielectric function and, therefore, does not have its charge absorbed by the background geometry. The remaining Maxwell field, which interacts with the dielectric function and whose charge disappears from the metric solution, plays the role of an auxiliary field, acting to model the horizon function. The idea of using an additional Maxwell field to induce charged backgrounds when looking for analytical Lifshitz black holes was used in \cite{brynjolfsson2010holographic} to model superconducting phase transitions and then other charged solutions were found by using the same mechanism (for example, \cite{pang2010charged,dehghani2011charged,alvarez2014nonlinearly,zangeneh2015thermodynamics}). Here, we also explore analytical aspects of thermodynamics and its associated critical behavior under the extended phase space formalism, where the cosmological constant is used to define the black hole pressure and the mass of the black hole solution is related to its enthalpy \cite{kastor2009enthalpy,dolan2011cosmological,dolan2011pressure,kubizvnak2012p,kubizvnak2017black}. Due to the number of parameters involved and the need for extensive numerical treatment, we decided to focus here on the analytical aspects of thermodynamics and, in the future, extend this study in a subsequent work.

This work is organized as follows. In Sec. II we present the framework of the model, discussing the role of the general covariance breaking in Einstein's equations and presenting its compatibility constraint. In Sec. III we present the black hole solution and explore some of its properties, such as conserved charges and extremal regime. In Sec. IV we study its thermodynamics and in Sec. V we analyze its critical behavior from the extended thermodynamics formalism. We conclude with final comments and future perspectives on the subject.
\section{Framework}
In this work we are interested in studying systems modeled by the effective action
\begin{eqnarray}\label{action}
S&=&\int d^{D} x\sqrt{-g}\left(\frac{1}{2}R-\Lambda-\frac{1}{2}\nabla_a \phi\nabla^a\phi-V(x,\phi)-\frac{1}{2}F_{ab}F^{ab}-\frac{1}{2}\varepsilon(x)\mathcal{F}_{ab}\mathcal{F}^{ab}\right),
\end{eqnarray}
where $g=\text{det}(g_{ab})$ denotes the metric determinant, $R$ is the Ricci scalar, $\Lambda=-\left(D-2\right)\left(D+3z-4\right)/2\ell^2$ denotes a negative cosmological constant and $F_{ab}=\nabla_a A_b-\nabla_b A_a$ is the usual Maxwell tensor. The scalar field $\phi(x)$ self-interacts through the scalar potential $V(x,\phi)$, which explicitly depends on the background coordinates $x^a$, with $a=0,1,\cdots,D-1$ and we also equip the model with a ``Maxwell-like'' tensor $\mathcal{F}_{ab}=\nabla_a \mathcal{B}_b-\nabla_b \mathcal{B}_a$ governing the dynamics of an auxiliary vector field $\mathcal{B}_a$ which mimics the Maxwell field and interacts with a nondynamical dielectric $\varepsilon(x)$. The field equations derived from the action \eqref{action} are
\begin{subequations}\label{ee}
\begin{eqnarray}
\label{ee1}\Box \phi-\frac{\partial V}{\partial\phi}&=&0,\\[1pt]
\label{ee2}\nabla_a F^{ab}&=&0,\\[3pt]
\label{ee3}\nabla_a \left(\varepsilon(x) \mathcal{F}^{ab}\right)&=&0,\\[3pt]
\label{ee4}\mathcal{E}_{ab}=G_{ab}+\Lambda g_{ab}-T_{ab}&=&0,
\end{eqnarray}
\end{subequations}
where $\Box=g^{ab}\nabla_a\nabla_b$ denotes the d'Alembertian operator, $G_{ab}=R_{ab}-g_{ab}R/2$ is the Einstein tensor, $\mathcal{E}_{ab}$ denotes a ``zero tensor" - useful for calculations -   and
\begin{eqnarray}\label{emt1}
T_{ab}&=&\nabla_a\phi\nabla_b\phi-\frac{1}{2}g_{ab}\left(\nabla\phi\right)^2-g_{ab}V(x,\phi)+2F_{a}^{~c}F_{bc}-\frac{1}{2}g_{ab}F_{cd}F^{cd}+\varepsilon(x)\left(2\mathcal{F}_{a}^{~c}\mathcal{F}_{bc}-\frac{1}{2}g_{ab}\mathcal{F}_{cd}\mathcal{F}^{cd}\right),~~~~~ 
\end{eqnarray}
represents the energy-momentum tensor. 

The explicit coordinate dependence in the action \eqref{action} is read as the presence of nondynamical background fields which breaks diffeomorphism invariance. It implies that $T_{ab}$ is no longer conserved, which induces a compatibility problem in the Einstein equation \eqref{ee4} since the contracted Bianchi identity remains valid for the tensor $G_{ab}$ (i.e., here we have $\nabla_a T^a_{~b}\neq0$ but $\nabla_a G^a_{~b}=0$ always). Therefore, one can only capture effective solutions for cases where the sums of contributions arising from the nondynamical background fields combine in such a way that $\nabla_a T^a_{~b}=0$, which leads us to the compatibility condition
\begin{equation}\label{boundeq}
    \partial_a V(x, \phi)=-\frac{1}{2}\mathcal{F}_{bc}\mathcal{F}^{bc}\partial_a \varepsilon(x),
\end{equation}
acting as a constraint on the field equations \eqref{ee}. The auxiliary field $\mathcal{B}_a$ mimics the Maxwell field $A_a$ and thus contributes to the background geometry with a free parameter which mimics electric charge. However, as shown below, by using the compatibility equation \eqref{boundeq} within the field equations \eqref{ee4} one can eliminate from the background solution the explicit dependence on this new conserved charge. Thus, the Maxwell field $A_a$ becomes the only responsible for providing charged contributions to the background geometry.
\section{Black hole solution}
The existence of well-behaved minimal energy soliton-like solutions as probe fields on asymptotically Lifshitz spacetimes \cite{moreira2022analytical,moreira2022scalar} gives rise to natural questions about how the insertion of backreactions into the model affects the structure of the fields and how the presence of dynamical and nondynamical field contributions shapes background geometries whose metric has a generic structure given by
\begin{equation}\label{backmetric}
ds^2=-\left(\frac{r}{\ell}\right)^{2z}e^{2\nu(r)}dt^2+\left(\frac{\ell}{r}\right)^{2}\frac{dr^2}{e^{2\nu(r)}}+\left(\frac{r}{\ell}\right)^{2}\hat{\sigma}_{ij}dx^i dx^j,
\end{equation}
where $z$ denotes the dynamical exponent, $\ell$ is a length scale, $(x^0,x^1)=(t,r)$ and $i,j=2,\cdots, D-1$. We assume that the horizon metric $\hat{\sigma}_{ij}(x^k)$ depends on the coordinates $x^i$'s, $2\leq i\leq D-1$, and describes a closed transverse $(D-2)$-dimensional Einstein manifold $\hat{\Sigma}_{\gamma}$ whose  Ricci tensor is $\hat{R}_{ij}=(D-3)\gamma\hat{\sigma}_{ij}$, with $\gamma=0,\pm1$. Thus, the manifold $\hat{\Sigma}_{\gamma}$ can hold spherical, planar or hyperbolic topologies for $\gamma=1$, $\gamma=0$ or $\gamma=-1$, respectively. The standard Lifshitz background \eqref{lifmetric} is retrieved in the limit $\nu(r)\to 0$ for $\hat{\sigma}_{ij}=\delta_{ij}$ and for $z=1$ we are led to AdS$_{D}$ setups. We also demand the scalar field, the vector fields and the effective dielectric to only have radial dependence, i.e.,
\begin{eqnarray}
\phi=\phi(r), ~A=A_adx^a=A(r)dt,~\mathcal{B}=\mathcal{B}_adx^a=B(r)dt~\text{and}~\varepsilon(x)=\varepsilon(r).~~~~~~
\end{eqnarray}
Under these requirements, Eq.\eqref{ee2} and Eq.\eqref{ee3} becomes
\begin{equation}\label{vf}
    A'(r)=\frac{q}{\ell}\left(\frac{\ell}{r}\right)^{D-z-1},~~B'(r)=\frac{\widetilde{q}/\ell}{\varepsilon(r)}\left(\frac{\ell}{r}\right)^{D-z-1},
\end{equation}
respectively, where $\left(q,\widetilde{q}\right)$ denotes integration constants and prime denotes derivation in relation to the $r$-coordinate. Note that here the solution for the Maxwell field can be obtained directly from a simple integration, but obtaining a solution for the auxiliary field $\mathcal{B}_a$ depends on the expression of the effective dielectric $\varepsilon(r)$ which, in turn, cannot be arbitrarily chosen due to its role in the constraint \eqref{boundeq}. The conserved charges associated to the Maxwell and the auxiliary field $\mathcal{B}_a$ are
\begin{subequations}
\begin{eqnarray}
Q&=&-\frac{1}{4\pi}\oint_{\partial\Sigma} d^{\small{D-2}}x \sqrt{|h^{(2)}|} n_a s_b F^{ab}=\frac{\omega_{D-2}^{(\gamma)}}{4\pi\ell}q,\\[3pt]
\widetilde{Q}&=&-\frac{1}{4\pi}\oint_{\partial\Sigma} d^{\small{D-2}}x \sqrt{|h^{(2)}|} n_a s_b \varepsilon (x)\mathcal{F}^{ab}=\frac{\omega_{D-2}^{(\gamma)}}{4\pi\ell}\widetilde{q},~~~~~~~~
\end{eqnarray}
\end{subequations}
respectively, where $n_a$ and $s_a$ are timelike and spacelike unit normal vectors to the  surface $\partial\Sigma$ defined at fixed $(r,t)$, equipped with the induced metric $h_{ij}^{(2)}=\left(\frac{r}{\ell}\right)^2\hat{\sigma}_{ij}$, and $\omega_{D-2}^{(\gamma)}=\oint_{\Hat{\Sigma}_\gamma}d^{D-2}x\sqrt{|\Hat{\sigma}_\gamma|}$ denotes the volume of the closed surface $\hat{\Sigma}_{\gamma}$.

We also set $V(x,\phi)=V(r,\phi)$ for the scalar potential and in this way the $(r,r)$ and $(t,t)$ components of $\mathcal{E}^a_{~b}=0$ in Eq.\eqref{ee3} becomes
\begin{subequations}\label{eeq}
\begin{eqnarray}
\label{err} \mathcal{E}^{r}_{~r}&=&\frac{(D-2)/2\ell^2}{r^{D+2z-4}}\left(r^{D+2z-3}e^{2\nu}\right)'-\frac{1}{2}\left(\frac{r}{\ell}\right)^{2}\phi'^2 e^{2\nu}-\frac{\hat{\gamma}}{2r^2}+\Lambda+\frac{q^2}{\ell^2}\left(\frac{\ell}{r}\right)^{2(D-2)}+\frac{\widetilde{q}^2/\ell^2}{\varepsilon(r)}\left(\frac{\ell}{r}\right)^{2(D-2)}+V,\label{eer}\\[8pt]
\label{ett}\mathcal{E}^{t}_{~t}&=&\frac{(D-2)/2\ell^2}{r^{D-2}}\left(r^{D-1}e^{2\nu}\right)'+\frac{1}{2}\left(\frac{r}{\ell}\right)^{2}\phi'^2 e^{2\nu}-\frac{\hat{\gamma}}{2r^2}+\Lambda+\frac{q^2}{\ell^2}\left(\frac{\ell}{r}\right)^{2(D-2)}+\frac{\widetilde{q}^2/\ell^2}{\varepsilon(r)}\left(\frac{\ell}{r}\right)^{2(D-2)}+V,\label{eet}
\end{eqnarray}
\end{subequations}
with $\hat{\gamma}=(D-2)(D-3)\gamma/\ell^2$. The remaining equations $\mathcal{E}^i_{~j}=0$ are left for checking. By dealing with the difference $\mathcal{E}^{r}_{~r}-\mathcal{E}^{t}_{~t}=0$ one can find the identity
\begin{equation}\label{phieq}
    \left(\frac{d\phi}{dr}\right)^2=\frac{(D-2)(z-1)}{r^2},
\end{equation}
whose solution is valid for $z\geq1$ and presented below along with the solution we found for the Maxwell field from Eq.\eqref{vf},
\begin{subequations}
\begin{eqnarray}
     \label{sfs}   \phi(r)&=&\varphi\pm\sqrt{(D-2)(z-1)}\ln(r/\ell),\\[3pt]
  A(r)&=&\Phi-\frac{q}{D-z-2}\left(\frac{\ell}{r}\right)^{D-z-2},
    \end{eqnarray}
\end{subequations}
where $\left(\varphi,\Phi\right)$ are integration constants\footnote{Here, we conveniently express the integration constant in the Maxwell field to coincide with the electrostatic potential at the horizon, as will become clear later in the thermodynamic analysis of the solution.}.  

The first consequence of inserting backreaction in the model occurs in the scalar field, which here acquires a logarithmic behavior. We also need to solve the equations \eqref{err} and \eqref{ett} to determine the background geometry and therefore it is necessary to choose an appropriate interaction potential, which must also provide a path to obtain the dielectric function $\varepsilon (r)$. In order to keep the control on the field equations and motivated by the procedure established in \cite{moreira2022scalar}, we chose to write the scalar potential as 
\begin{equation}\label{1potmodel}
V(r,\phi)=\frac{1}{2}e^{-2\nu}\left(\frac{\ell}{r}\right)^{2(D+z-2)}\left(\frac{dW}{d\phi}\right)^{2}+U(r),
\end{equation}
where $W=W\left(\phi\right)$ and $U(r)$ are auxiliary functions to be specified from the field equations. In the probe regime, the auxiliary function $W(\phi)$ models the scalar field solution and provides its boundary values, but here it acts on the first sector of the potential to describe the self-interaction of the scalar field together with an explicitly coordinate-dependent factor arranged to be compatible with the second-order scalar field equation \eqref{ee1}, which now becomes
\begin{equation}\label{1ordereq}
\frac{d\phi}{dr}=\pm  \left(\frac{\ell}{r}\right)^{D+z-1}\frac{dW}{d\phi}e^{-2\nu},
\end{equation}
which is first-order and allows one to find the expression for $W(\phi)$, since the scalar field solution \eqref{sfs} is invertible. Unfortunately, the expression we found  for $W(\phi)$ is awkward so we decided to omit it here. The auxiliary function $U(r)$ acts as a nondynamical field which permeates the system, but does not interact with the scalar and vector fields.

By using  Eq.\eqref{phieq}, Eq.\eqref{1potmodel} and Eq.\eqref{1ordereq} one can show that the pair of functions 
\begin{subequations}
\begin{eqnarray}
\label{dielfunc}   \frac{1}{\varepsilon(r)}&=&\frac{\left(z-1\right)\left(D+z-2\right)}{2\widetilde{q}^2}\left(\frac{r}{\ell}\right)^{2(D-2)}e^{2\nu},~~~~\\[2pt]
 \label{ur}   U(r)&=&-\frac{z\left(z-1\right)}{2\ell^2}e^{2\nu},\label{ufunc}
\end{eqnarray}
\end{subequations}
satisfies the compatibility equation \eqref{boundeq}. One can also note that both expressions above approach zero in the limit $z\to 1$ and that Eq.\eqref{dielfunc} can be rewritten as 
\begin{equation}
    \frac{\widetilde{q}^2/\ell^2}{\varepsilon(r)}\left(\frac{\ell}{r}\right)^{2(D-2)}=\frac{(z-1)(D+z-2)}{2\ell^2}e^{2\nu}.
\end{equation}
By inserting the above expression in the set of equations presented in Eq.(\ref{eeq}), we eliminate from the background geometry  its dependence on the free parameter related to the auxiliary field $\mathcal{B}_a$. In this way, the contributions associated with the general covariance breaking ingredients are distributed in different sectors of the field equations and we can finally find as solution
\begin{eqnarray}\label{metricsol}
e^{2\nu(r)}=1&-&2m\left(\frac{\ell}{r}\right)^{D+3z-4}+\frac{\gamma_z}{r^2}+q^2_z\left(\frac{\ell}{r}\right)^{2(D-2)},~~~~~~
\end{eqnarray}
where, for simplicity, we define the constants
\begin{equation}
    \gamma_z=\frac{D-3}{D+3z-6}\gamma~~\text{and}~~q^2_z=\frac{2 q^2}{\left(D-2\right)\left(D-3z\right)}. 
\end{equation}
The solution \eqref{metricsol} is well behaved for $D\neq 3z$\footnote{One could regularize the solution \eqref{metricsol} for $D=3z$ by applying a simple rescaling on the charge parameter $q$ when defining it in Eq.\eqref{vf}. However, depending on how this step is performed, pathologies may appear in the Maxwell field for $D-3z<0$ or we find $Q=0$ for $D=3z$, retrieving the solution already found in \cite{moreira2022scalar}. In this way, we chose to keep the background solution undefined for this case.} and in order to achieve Lifshitz spacetime in the asymptotic regime $r\to\infty$ we must set $D+3z-4>0$. 

\begin{figure}[t!]
    \begin{center}
\begin{tabular}{cc}
\begin{minipage}{0.5\textwidth} 
\hspace{-1cm}\includegraphics[width=0.75\textwidth]{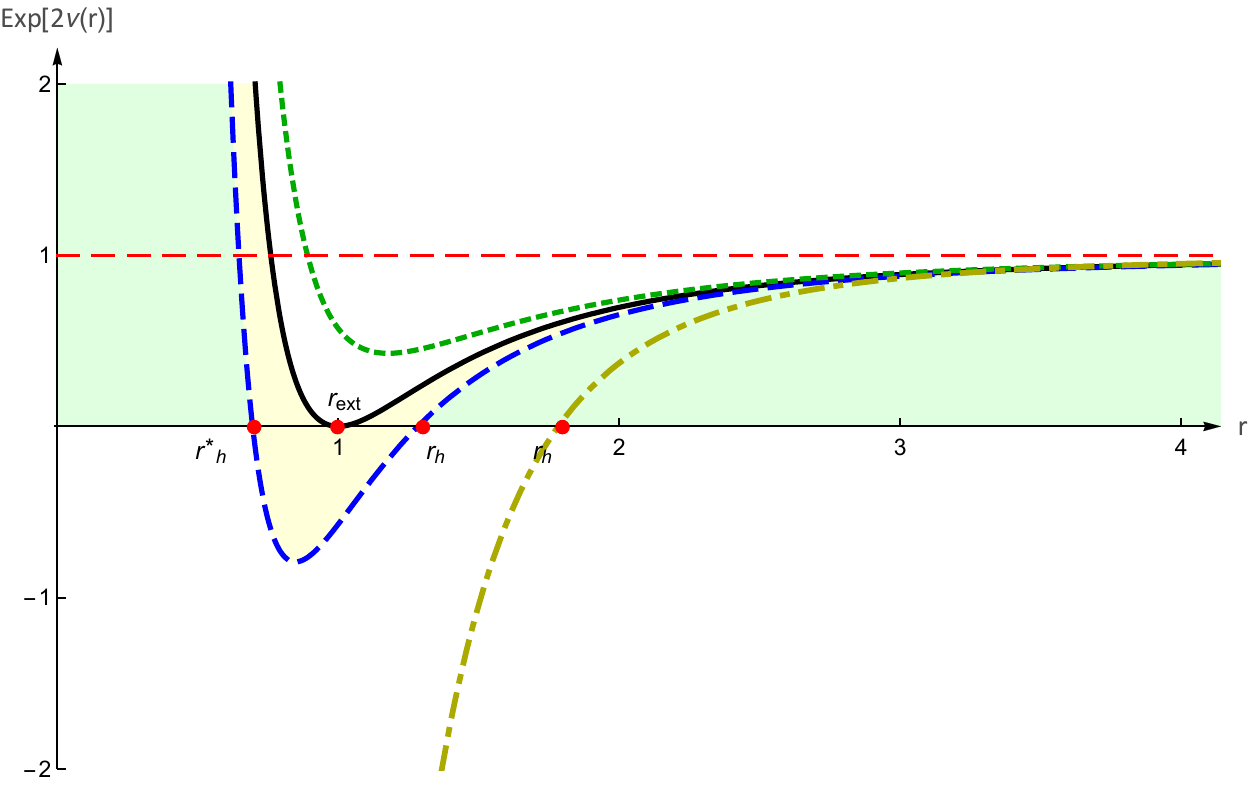}\\
\hspace{-1cm}\small (a)
\\ 
\hspace{-1cm}\includegraphics[width=0.75\textwidth]{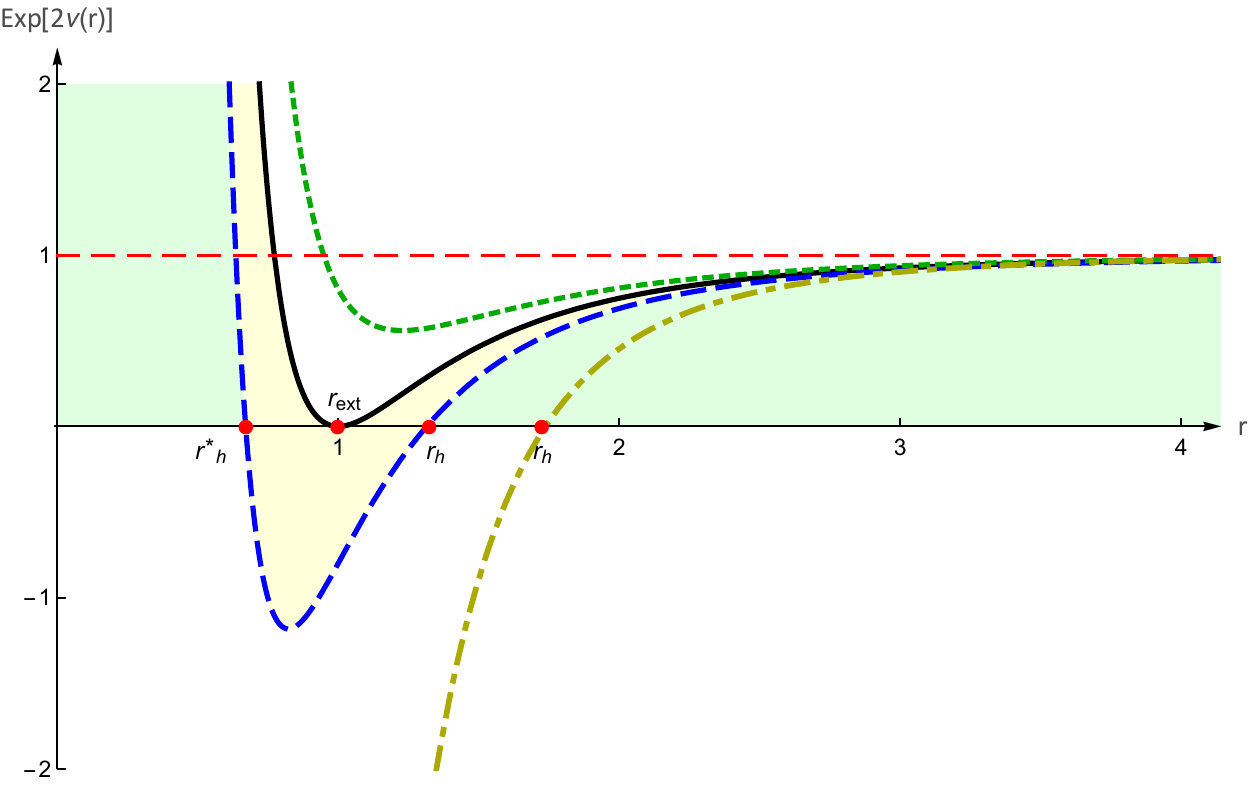}\\
\hspace{-1cm}\small (b)
\end{minipage}
\begin{minipage}{0.5\textwidth} 
\hspace{-1cm}
\includegraphics[width=0.95\textwidth]{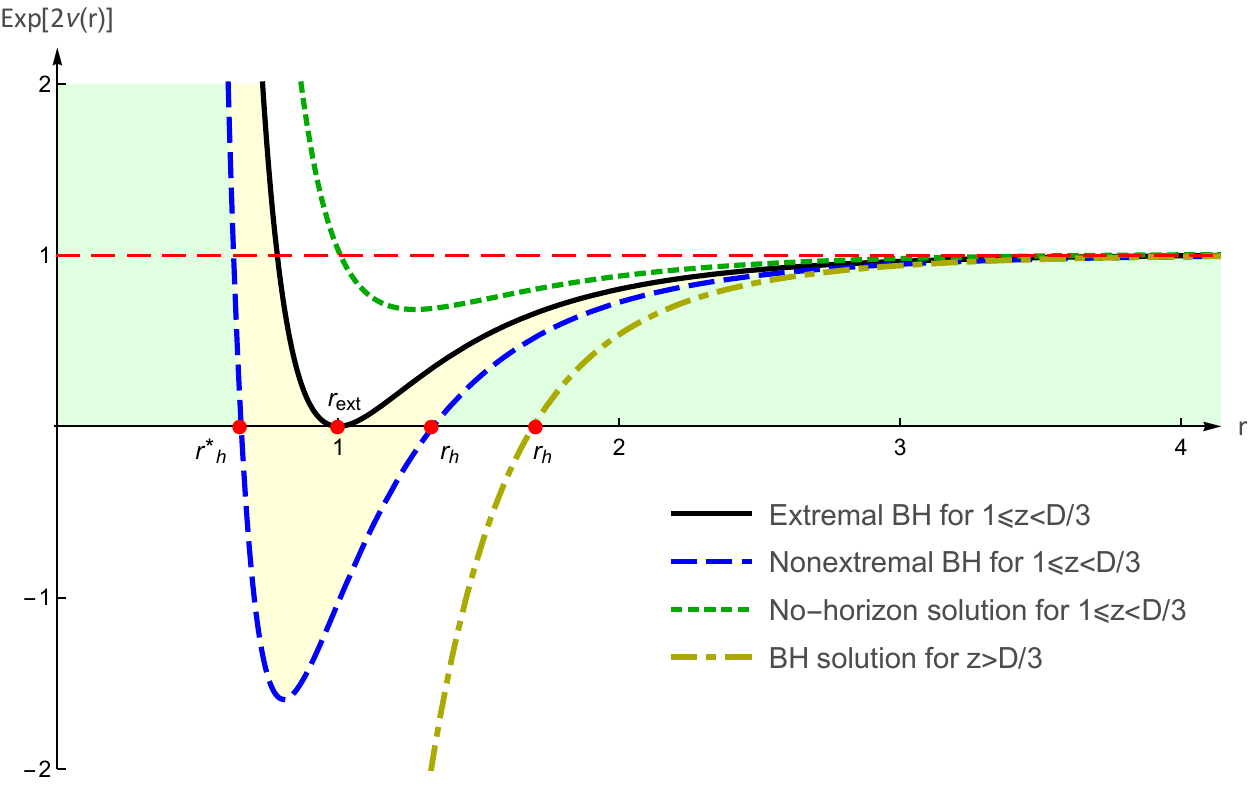}\\
\hspace{-1cm}\small (c)\end{minipage}
\end{tabular}
\caption{Background solution \eqref{backmetric} for different parameter values. We used here $D=4, \ell=10$ and (a) $\gamma=-1$, (b) $\gamma=0$ and (c) $\gamma=1$. For the cases where $1\leq z<D/3$ we used $z=5/4$ and $r_{ext}=1$. In the blue curve (dashed) denoting a nonextreme black hole we use a mass in excess of 5\% above the extremal mass and in the green curve (dotted, no horizon) we used a mass 5\% less. For the case where $z>D/3$ (yellow, dot-dashed curve) we used $z=5/3$, $m=2.5\times 10^{-5}$ and $q=0.025$.}\label{fig1}
    \end{center}
    \end{figure}

The existence of a horizon for some radius $r = r_h>0$ depends on the values of the parameters involved in the solution. In particular, one can find horizons for $D-3z<0$ since in these cases the ``mass parameter'' term dominates the solution \eqref{metricsol} in the limit $r\to 0$ for $m>0$, which implies that $e^{2\nu(r\to0)}\to-\infty$ and thus there is $r=r_h$ such that $e^{2\nu(r_h)}=0$. For $D-3z>0$, however, the ``charge parameter'' term dominates the background in the limit $r\to 0$ and leads us to a divergence as $e^{2\nu(r\to0)}\to+\infty$, which does not ensure the existence of a horizon. In fact, studying the extreme values of the function $e^{2\nu(r)}$ we observe that it must have at least one event horizon if there is an absolute minimum for a given radius $r=r_0$ such that the inequality
\begin{equation}\label{ineq}
     1+(D+3z-6)m\left(\frac{\ell}{r_0}\right)^{D+3z-4}\leq(D- 3)q_z^2\left(\frac{\ell}{r_0}\right)^{2(D-2)},
\end{equation}
is satisfied. This bound is saturated by extremal (zero temperature) black hole solutions, whose mass and charge parameters are given by
\begin{subequations}
\begin{eqnarray}
    m_{ext}&=&\frac{D-2}{D-3z}\left(\frac{r_{ext}}{\ell}\right)^{D+3z-4}\left(1+\frac{D-3}{D-2}\frac{\gamma_z}{r^2_{ext}}\right),~~~~~~~~~\\[3pt]
\nonumber q^2_{ext}&=&\frac{\hat{\gamma}}{2}\left(\frac{r_{ext}}{\ell}\right)^{2(D-3)} + \frac{(D-2)(D+3z-4)}{2}\left(\frac{r_{ext}}{\ell}\right)^{2(D-2)},~~~~~~~
\end{eqnarray}
\end{subequations}
respectively, where $r=r_{ext}$ denotes the extremal radius. Respecting the conditions for the existence of the event horizon, we can write
\begin{eqnarray}
    m(r_h)=\frac{1}{2}\left(\frac{r_h}{\ell}\right)^{D+3z-4}\left(1+\frac{\gamma_z}{r_h^2}+q_z^2\left(\frac{\ell}{r_h}\right)^{2(D-2)}\right),~~~~~~~~
\end{eqnarray}
and studying its extreme values, we observe that $m_{ext}$ denotes a lower bound for the mass parameter values, i.e., $m\geq m_{ext}$. It is also worth mentioning that for black holes with planar horizons the minima of the solution \eqref{metricsol} occurs at
\begin{eqnarray}
r_{0}=\left(\frac{2q^2/m}{(D-3z)(D+3z-4)}\right)^{\frac{1}{D-3z}}\ell,~~ \left(\text{for}~\gamma=0\right)~~~~~~~
\end{eqnarray}
for any allowed values of the parameters $\left(q^2,m\right)$ and for black holes with hyperbolical horizons we must have 
\begin{eqnarray}\label{brext}
 r_{ext}\geq \sqrt{\frac{D-3}{D+3z-4}}~~\left(\text{for}~~\gamma=-1\right).
\end{eqnarray}
in order to keep $q^2\geq0$. The typical behavior of the background solution \eqref{backmetric} for different parameter values is depicted in Fig.\eqref{fig1} and illustrate how the causal structure of the background geometry is affected by both the presence of the fields and the presence of terms violating diffeomorphism invariance. One can note that for cases where $z>D/3$ there is a single event horizon while for cases where $D/3>z\geq1 $ there is an inner Cauchy horizon $r_h^*$ and an outer event horizon $r_h$ whose values approach as the black hole mass parameter approaches $m_{ext}$, becoming degenerate at the extremal limit. For $z=1$, general covariance is restored and the scalar field becomes trivial, thus we recover the Reissner-Nordstrom-AdS black hole, whose nature of the Cauchy horizon and its instability is explored in \cite{hartnoll2020gravitational,hartnoll2021diving}. The emergence of Cauchy horizons for $D/3>z\geq1 $ indicates that the loss of predictability in the inner region of the event horizon which occurs in the covariant regime persists when terms violating diffeomorphism invariance are related to small spacetime anisotropies but ceases when the deviations from the relativistic regime are high enough $(z>D/3)$, since in these cases the Cauchy horizon is removed. It is also worth pointing out that there are some ``no inner horizon'' theorems for an extensive class of Einstein-Maxwell-scalar models \cite{hartnoll2020gravitational,hartnoll2021diving,cai2021no,an2021no} where the presence of neutral and charged scalar fields makes impossible the presence of Cauchy horizons in a large set of black hole families - including some Lifshitz black holes - but the diffeomorphism invariance breaking imposed on the model action allows us to evade them.

\begin{figure}[ht!]
  \centering
  \begin{tabular}{ c @{\quad} c }
    \includegraphics[scale=0.4]{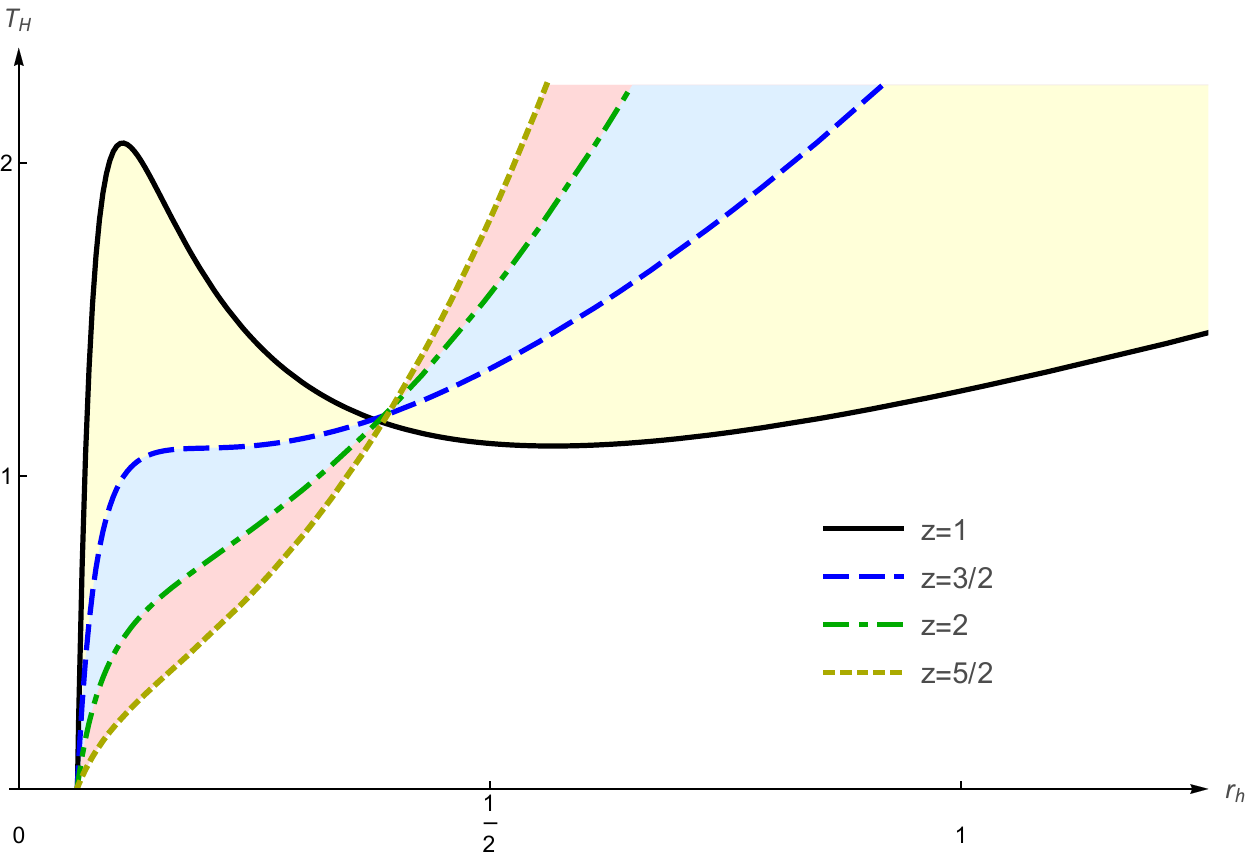} &
      \includegraphics[scale=0.4]{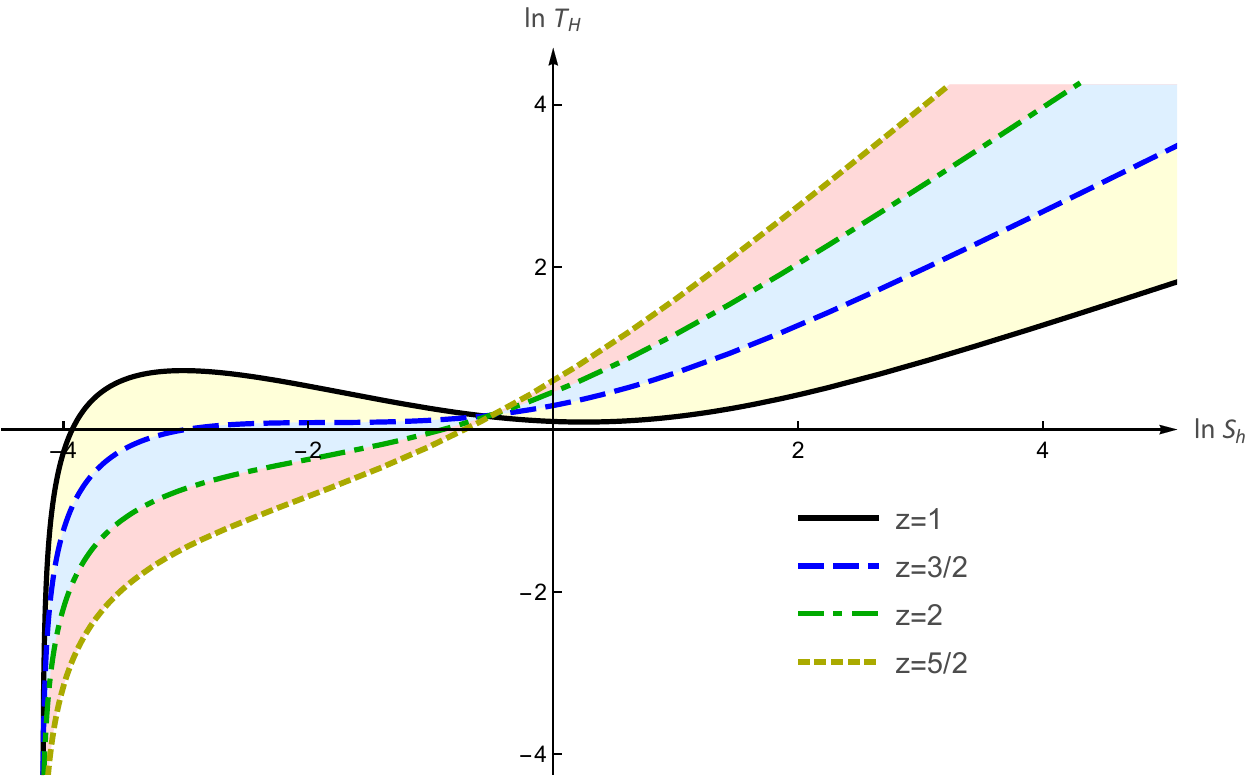} \\
    \small (a) &
      \small (b)
  \end{tabular}
  \vspace*{8pt}
  \caption{Temperature as a function of (a) the event horizon and (b)  the entropy for $D=4$,  $\ell=0.5$  $q=0.25$, $\gamma=1$ and some different values of the $z$.}\label{figtemp}
\end{figure}

\section{Thermodynamics}

Here we study the thermodynamics of the solution found from the point of view of the extended thermodynamics \cite{kastor2009enthalpy,dolan2011cosmological,dolan2011pressure,kubizvnak2017black}, where the pressure is defined as $P=-\Lambda/8\pi $ and the mass of the black hole is identified as the enthalpy. Due to the number of parameters and the nonlinearities involved, which require extensive numerical treatment, we have decided to present, for now, a discussion limited to the analytical aspects of thermodynamics and, in the future, to develop a complementary work aimed at deepening and extending this topic.

The temperature and entropy associated with the black hole solution \eqref{metricsol} are
\begin{subequations}
\begin{eqnarray}\label{bht}
    T_{H}&=&\frac{r_h^z}{4\pi \ell^{z+1}}\left(c_1+\frac{(D-3)\gamma}{r_h^2}-\frac{2q^2}{D-2}\frac{\ell^{2(D-2)}}{r_h^{2(D-2)}}\right),~~~~~~~\\[3pt]
    S_{bh}&=&\frac{1}{4}\left(\frac{r_h}{\ell}\right)^{D-2} \omega^{\left(\gamma\right)}_{D-2},
\end{eqnarray}
\end{subequations}
respectively, where $c_1=D+3z-4$. For $\left(\gamma, q, z\right)=\left(0, 0, 1\right)$ the temperature reproduces the typical linear behavior of black strings, as expected \cite{lemos1995three,darlla2023black}. The temperature behavior as a function of the event horizon and entropy for different values of the critical exponent $z$ are depicted in Fig. (\ref{figtemp}a) and Fig. (\ref{figtemp}b), respectively. In particular, by setting $T_H=0$ for $D=4$ one can find for the extremal radius 
\begin{eqnarray}    \left.r_{ext}\right|_{D=4}=\sqrt{\frac{\sqrt{\gamma^2+12 q^2 \ell^4 z}-\gamma}{6z}},
\end{eqnarray}
which is zero for $q=0$, $\gamma=0,1$ and saturates the bound \eqref{brext} for $q=0$, $\gamma=-1$, as expected.

Since we are working on charged backgrounds, the first law of thermodynamics becomes
\begin{eqnarray}
\label{firstlaw}dM&=&T_{H}dS_{bh}+V_{bh}dP+\Phi dQ,
\end{eqnarray}
where $V_{bh}$ represents the thermodynamic volume, $Q$ is the electric charge and 
\begin{equation}\label{elecpot}
    \Phi=\frac{q}{D-z-2}\left(\frac{\ell}{r_h}\right)^{D-z-2}, 
\end{equation}
which sets $A(r_h)=0$ and denotes the electrostatic potential on the
horizon, which diverges for $z=D-2$ in cases where $q\neq0$.  The thermodynamical mass calculated from integrating the first law \eqref{firstlaw} from zero to the horizon at constant pressure and charge is given by
\begin{eqnarray}
\nonumber M&=&\frac{\omega_{D-2}^{(\gamma)}}{16\pi\ell}\left(\frac{\hat{\gamma}}{D+z-4}\left(\frac{r_h}{\ell}\right)^{D+z-4}+\frac{16\pi\ell^2 P}{D+z-2}\left(\frac{r_h}{\ell}\right)^{D+z-2}+\frac{2q^2}{D-z-2}\left(\frac{\ell}{r_h}\right)^{D-z-2}\right),
\end{eqnarray}
which is identified with the enthalpy of the system \cite{kastor2009enthalpy}. In this way, the black hole volume becomes
\begin{eqnarray}
V_{bh}=\left.\frac{\partial M}{\partial P}\right|_{S_{bh},Q}=\frac{\ell}{D+z-2}\left(\frac{r_h}{\ell}\right)^{D+z-2}\omega_{D-2}^{(\gamma)}, ~~~~~
\end{eqnarray}
and a direct check reveals that the electrostatic potential \eqref{elecpot} can also be calculated from the relation
\begin{equation}
    \Phi=\left.\frac{\partial M}{\partial Q}\right|_{\left(V_{bh},T_H\right)},
\end{equation}
ensuring the internal consistency of the  expressions found so far. Finally, one can use the expressions above to deduce the Smarr-like relation
\begin{eqnarray}
\label{smarr}M=\frac{(D-2)T_H S_{bh}}{D+z-4}-\frac{2PV_{bh}}{D+z-4}+\frac{(D-3)\Phi Q}{D+z-4},~~~~~
\end{eqnarray}
which retrieves the Smarr relation of charged $AdS_{(D)}$ black holes for $z=1$ \cite{gunasekaran2012extended}.

\section{Critical behavior}
\begin{figure}[t!]
  \centering
  \begin{tabular}{ c @{\quad} c }
    \includegraphics[scale=0.4]{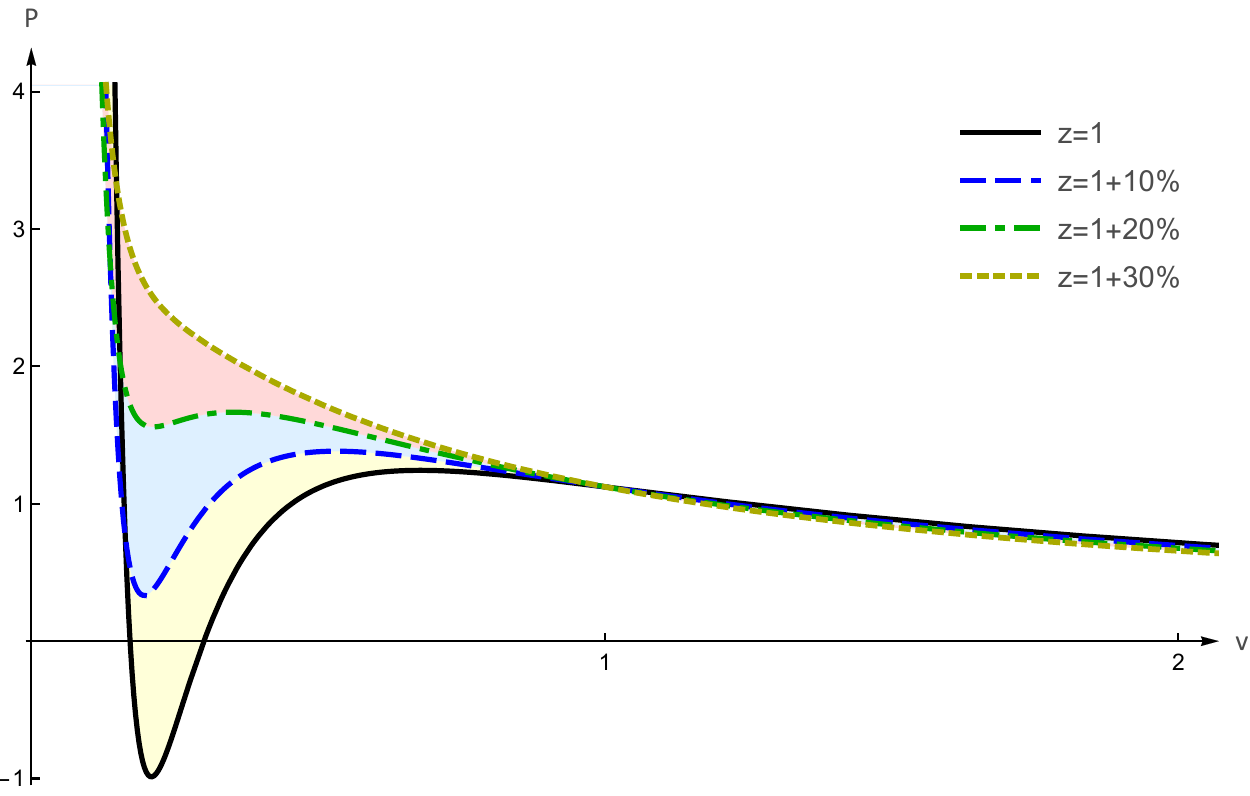} &
      \includegraphics[scale=0.4]{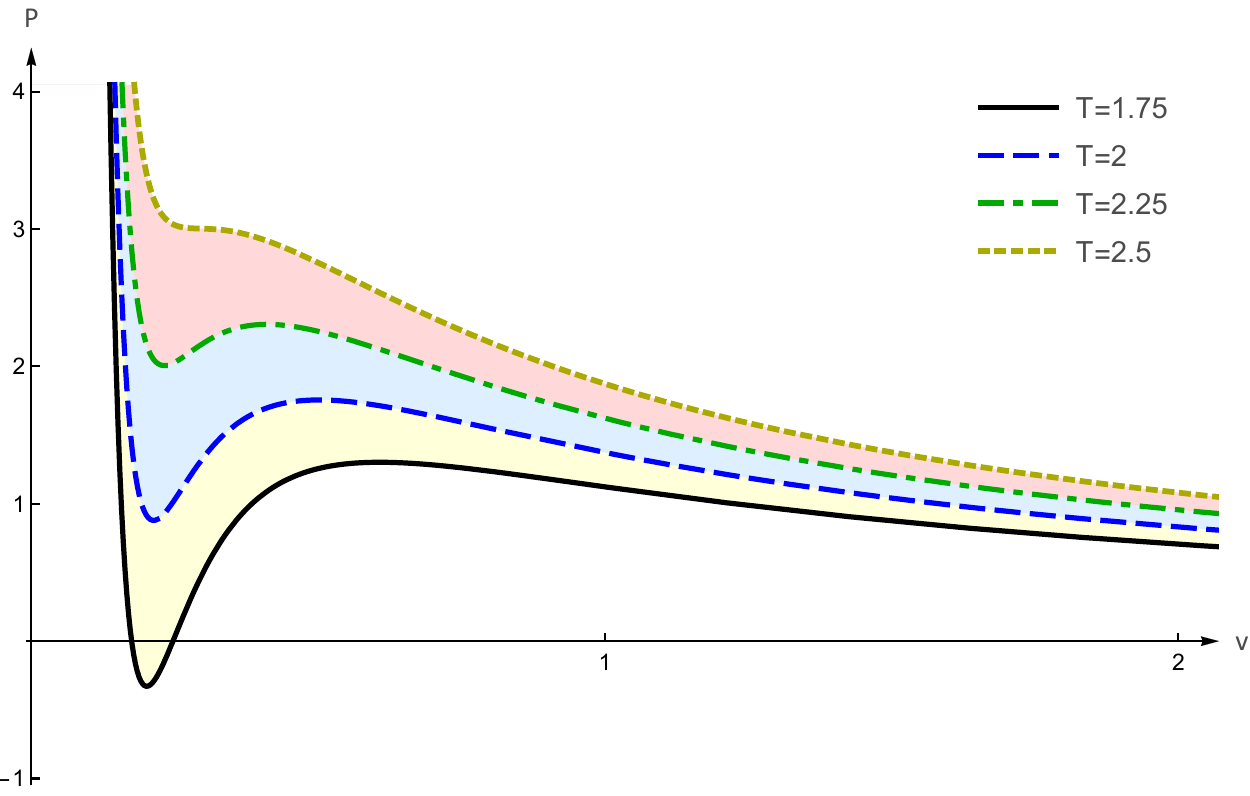} \\
    \small (a) &
      \small (b)
  \end{tabular}
  \vspace*{8pt}
  \caption{$P-v$ diagram for (a) different values of $z$ and fixed temperature $T=1.75$ and (b) different temperatures and fixed $z=1+5\%$. We settled $D=4$,  $q=0.25$, $\ell =0.5$ and $\gamma=1$.}\label{fig3}
\end{figure}
From the thermodynamical quantities presented in the previous section, one can analyze the critical behavior associated with the black hole solution found. In particular, from Eq.\eqref{bht} one can write the equation of state
\begin{equation}\label{stateeq}
    P(T,\upsilon)=\frac{T}{\upsilon}-\frac{a}{\upsilon^{2/z}}+\frac{b}{\upsilon^{2(D-2)/z}}
\end{equation}
where $T=T_H$, $\upsilon=\frac{4\ell}{D-2}\left(\frac{r_h}{\ell}\right)^{z}$ and
\begin{subequations}
\begin{eqnarray}
    a=\frac{\hat{\gamma}}{\pi\left(4\ell\right)^2}\left(\frac{4\ell}{D-2}\right)^{2/z},~~
    b=\frac{2q^2}{\pi\left(4\ell\right)^2}\left(\frac{4\ell}{D-2}\right)^{2(D-2)/z}.
\end{eqnarray}
\end{subequations}
Equation \eqref{stateeq} resembles the Van der Waals equation with zero co-volume for $(z, b,\gamma)=(1,0,1)$ and an ideal gas for $(z, b, \gamma)=(1,0,0)$. Both the powers and coefficients of the equation of state depend on the anisotropic parameter $z$ and furthermore the sign of the interaction term now also depends on the topology of the black hole, assuming an attractive, noninteracting or repulsive character for $\gamma=1, 0,-1$, respectively. It implies that the critical behavior only occurs in setups presenting spherical topology and for this case the spinodal curve connecting the isotherms minima and maxima for values $T\leq T_c$, where $T_c$ denotes the critical temperature, is given by
\begin{equation}
    P_{sp}=\left(\frac{2}{z}-1\right)\frac{a}{\nu^{2/z}}-\left(\frac{2(D-2)}{z}-1\right)\frac{b}{\nu^{2(D-2)/z}}.
\end{equation}
 The $P-v$ diagram for different temperatures and different values of $z$ are depicted in Fig. (\ref{fig3}a) and Fig. (\ref{fig3}b), respectively. Note that for a fixed temperature there is no critical behavior for large enough scaling anisotropies.

\begin{figure}[t!]	
		\centering
\includegraphics[width=0.6\linewidth]{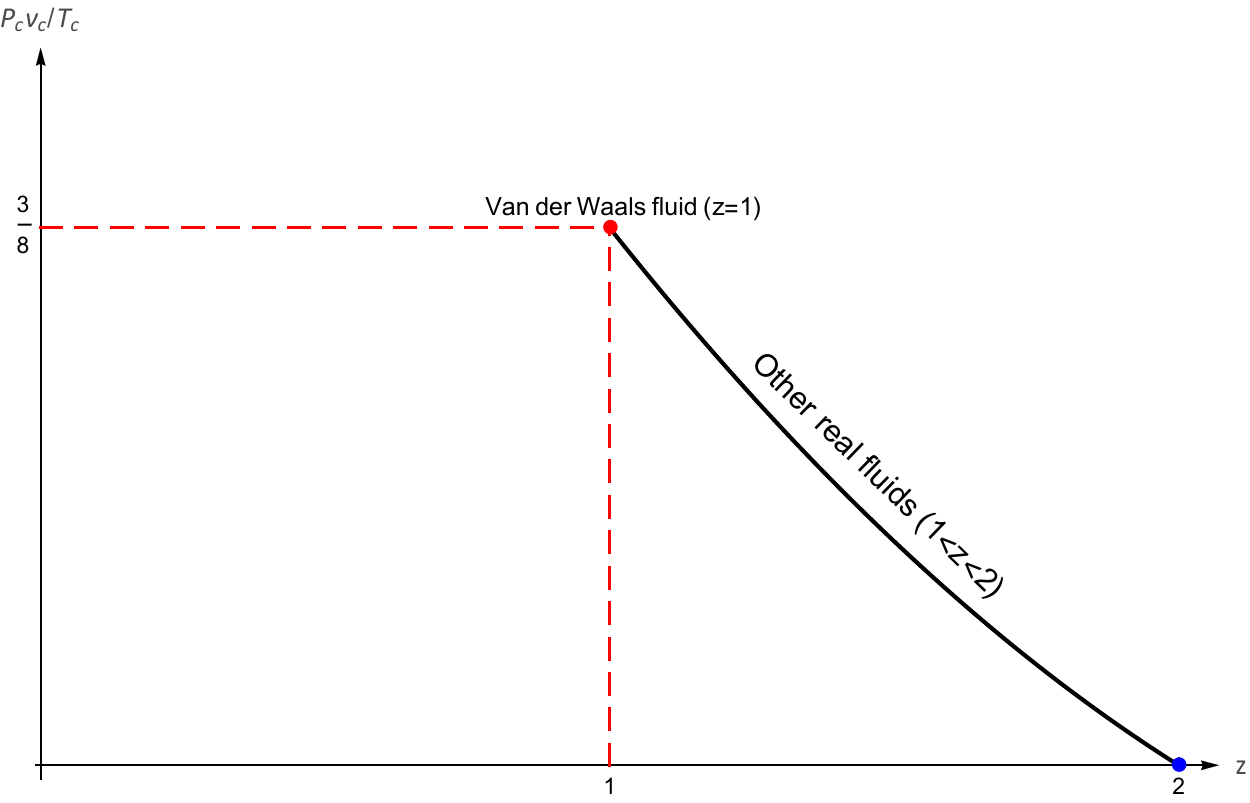}\\
\vspace{2mm}
\caption{Critical compressibility factor as a function of dynamical exponent for $D=4$.}
\label{fig4}
\end{figure}

We can determine the critical values of $P,v$ and $T$ from calculating the stationary point of inflection of the $P-v$ diagram, given by solving of the set of equations
\begin{equation}
    \left.\frac{\partial P}{\partial \upsilon}\right|_{T,Q}=0,~~~\left.\frac{\partial^2 P}{\partial \upsilon^2}\right|_{T,Q}=0.
\end{equation}
It lead us to the critical values
\begin{equation}
    \upsilon_c=\left(\frac{\left(D-2\right)\left(2D-4-z\right)}{2-z}\frac{b}{a}\right)^{\frac{z/2}{(D-3)}},~~~~T_c=\frac{4\left(D-3\right)}{z\left(2D-4-z\right)}\frac{a}{\upsilon_c^{(2-z)/z}} ~~~\text{and}~~~P_c=\frac{\left(D-3\right)\left(2-z\right)}{z\left(D-2\right)}\frac{a}{\upsilon_c^{2/z}},
\end{equation}
which are well-defined only for $1\leq z<2$. With these values, the $D$-dimensional critical compressibility factor becomes
\begin{equation}
   Z_c\equiv\frac{P_c\upsilon_c}{T_c}=\left(1-\frac{z}{2}\right)\left(1-\frac{z/2}{D-2}\right),
\end{equation}
which can be used to express the reduced equation of state as
\begin{equation}
     Z^{-1}_c\widetilde{T}= \widetilde{\nu}\left(\widetilde{P}+\frac{c_2}{\widetilde{\nu}^{2/z}}\right)-\frac{z/(D-3)}{2D-4-z}\widetilde{\nu}^{1-2(D-2)/z},
 \end{equation}
 where $c_2=z(D-2)/((2-z)(D-3))$, $\widetilde{\nu}=\nu/\nu_c$, $\widetilde{T}=T/T_c$ and $\widetilde{P}=P/P_c$. In particular, for $D=4$ one finds
\begin{equation}\label{compeq}
    \left.Z_c\right|_{D=4}=\frac{3}{8}-\delta
\end{equation}
where $\delta=(z-1)(5-z)/8$ and whose behavior is depicted in Fig.\eqref{fig4}. In the limit $z\to 2$ we have $Z_c\to 0$ and for $z=1$ the critical compressibility retrieves the value found for a charged AdS black hole, which coincides with the universal value predicted for the Van der Waals fluids. For $1<z<2$, however, one founds $3/8>Z_c>0$, which coincides with values experimentally measured for several real gases (see \cite{vargaftik1975tables} and related references). In this way, since the idealized Van der Waals model provides isotherms to approximately describe liquid-gas phase transitions of real fluids and the critical compressibility acts as a measure of how far a given fluid is from the Van der Waals fluid, Eq.\eqref{compeq} points to the possibility that Lifshitz black hole thermodynamics may be useful in effectively modeling real fluids beyond the Van der Waals regime.

\section{Ending comments} 
The study of global defect structures that evade the Derrick's theorem using general covariance breaking mechanisms has been put forward some time ago. Such structures were the first example where using explicitly coordinyate-dependent scalar potentials proved to be a nice alternative do find scalar global defects in arbitrary dimensions without adding gauge fields. In recent studies, the authors have extended this idea to curved spacetimes, focusing on obtaining global defects such as probe fields on fixed Lifshitz spacetimes, Lifhitz black holes and other types of black holes in the literature, where interesting new global solutions have been found. The studies at the probe-limit indicated how to advance in this topic to find complete field solutions in scenarios presenting backreaction, where the background geometry is assumed to be dynamical and a new topological Lifshitz black hole solution was found. The price to pay is the need of considering an auxiliary gauge field whose dependence is eliminated from the geometry by constraining the energy-momentum tensor to be divergenceless. We here extended these studies to find a charged Lifshitz black hole solution.  To ensure the charge of the black hole a second gauge field that correctly plays the role of the Maxwell field should also be added to the Lagrangian. We opted to approach black hole thermodynamics from the extended phase formalism and the presence of electric charge in the background geometry is important for the occurrence of critical behavior. The existence of anisotropic scaling provides rich thermodynamics, with interesting phase transitions whose thermal compressibility can be associated with real fluids other than the Van der Waals one for a given range of the dynamical exponent values. Since the new Lifshitz black hole geometry here was found from a method only recently explored, further studies on its stability and other aspects must be carried out. However, we believe that the mechanism presented here opens the way for several applications that goes from development of new topological solutions, new class of Lifshitz black holes and holographic description of several systems of interest such condensed matter and high energy physics whose phase transitions can be better understood in the anisotropic gravity dual.
\section*{Acknowledgments}
D.C.M. would like to thank the Brazilian agencies CNPq and FAPESQ-PB for partial financial support (PDCTR FAPESQ-PB/CNPq, Grant no. 317985/2021-3). A.S.L. acknowledges support from CAPES (Grant o. 88887.800922/2023-00) and F.A.B. acknowledges support from CNPq (Grant nos. 309092/2022-1) and also CNPq/PRO-
NEX/FAPESQ-PB (Grant nos. 165/2018), for partial financial support.
\bibliographystyle{elsarticle-num-names} 

  \bibliography{biblio} 

\begin{thebibliography}{87}
\expandafter\ifx\csname natexlab\endcsname\relax\def\natexlab#1{#1}\fi
\providecommand{\url}[1]{\texttt{#1}}
\providecommand{\href}[2]{#2}
\providecommand{\path}[1]{#1}
\providecommand{\DOIprefix}{doi:}
\providecommand{\ArXivprefix}{arXiv:}
\providecommand{\URLprefix}{URL: }
\providecommand{\Pubmedprefix}{pmid:}
\providecommand{\doi}[1]{\href{http://dx.doi.org/#1}{\path{#1}}}
\providecommand{\Pubmed}[1]{\href{pmid:#1}{\path{#1}}}
\providecommand{\bibinfo}[2]{#2}
\ifx\xfnm\relax \def\xfnm[#1]{\unskip,\space#1}\fi
\bibitem[{Maldacena(1999)}]{maldacena1999large}
\bibinfo{author}{J.~Maldacena},
\newblock \bibinfo{title}{The large-n limit of superconformal field theories
  and supergravity},
\newblock \bibinfo{journal}{International journal of theoretical physics}
  \bibinfo{volume}{38} (\bibinfo{year}{1999}) \bibinfo{pages}{1113--1133}.
\bibitem[{Aharony et~al.(2000)Aharony, Gubser, Maldacena, Ooguri, and
  Oz}]{aharony2000large}
\bibinfo{author}{O.~Aharony}, \bibinfo{author}{S.~S. Gubser},
  \bibinfo{author}{J.~Maldacena}, \bibinfo{author}{H.~Ooguri},
  \bibinfo{author}{Y.~Oz},
\newblock \bibinfo{title}{Large n field theories, string theory and gravity},
\newblock \bibinfo{journal}{Physics Reports} \bibinfo{volume}{323}
  (\bibinfo{year}{2000}) \bibinfo{pages}{183--386}.
\bibitem[{Ammon and Erdmenger(2015)}]{ammon2015gauge}
\bibinfo{author}{M.~Ammon}, \bibinfo{author}{J.~Erdmenger},
  \bibinfo{title}{Gauge/gravity duality: Foundations and applications},
  \bibinfo{publisher}{Cambridge University Press}, \bibinfo{year}{2015}.
\bibitem[{Hartnoll(2009)}]{hartnoll2009lectures}
\bibinfo{author}{S.~A. Hartnoll},
\newblock \bibinfo{title}{Lectures on holographic methods for condensed matter
  physics},
\newblock \bibinfo{journal}{Classical and Quantum Gravity} \bibinfo{volume}{26}
  (\bibinfo{year}{2009}) \bibinfo{pages}{224002}.
\bibitem[{Kachru et~al.(2008)Kachru, Liu, and Mulligan}]{kachru2008gravity}
\bibinfo{author}{S.~Kachru}, \bibinfo{author}{X.~Liu},
  \bibinfo{author}{M.~Mulligan},
\newblock \bibinfo{title}{Gravity duals of lifshitz-like fixed points},
\newblock \bibinfo{journal}{Physical Review D} \bibinfo{volume}{78}
  (\bibinfo{year}{2008}) \bibinfo{pages}{106005}.
\bibitem[{Taylor(2016)}]{taylor2016lifshitz}
\bibinfo{author}{M.~Taylor},
\newblock \bibinfo{title}{Lifshitz holography},
\newblock \bibinfo{journal}{Classical and Quantum Gravity} \bibinfo{volume}{33}
  (\bibinfo{year}{2016}) \bibinfo{pages}{033001}.
\bibitem[{Zaanen et~al.(2015)Zaanen, Liu, Sun, and
  Schalm}]{zaanen2015holographic}
\bibinfo{author}{J.~Zaanen}, \bibinfo{author}{Y.~Liu}, \bibinfo{author}{Y.-W.
  Sun}, \bibinfo{author}{K.~Schalm}, \bibinfo{title}{Holographic duality in
  condensed matter physics}, \bibinfo{publisher}{Cambridge University Press},
  \bibinfo{year}{2015}.
\bibitem[{Ho{\v{r}}ava(2009)}]{hovrava2009quantum}
\bibinfo{author}{P.~Ho{\v{r}}ava},
\newblock \bibinfo{title}{Quantum gravity at a lifshitz point},
\newblock \bibinfo{journal}{Physical Review D} \bibinfo{volume}{79}
  (\bibinfo{year}{2009}) \bibinfo{pages}{084008}.
\bibitem[{Sotiriou(2011)}]{sotiriou2011hovrava}
\bibinfo{author}{T.~P. Sotiriou},
\newblock \bibinfo{title}{Ho{\v{r}}ava-lifshitz gravity: a status report},
\newblock in: \bibinfo{booktitle}{Journal of Physics: Conference Series},
  volume \bibinfo{volume}{283}, \bibinfo{organization}{IOP Publishing},
  \bibinfo{year}{2011}, p. \bibinfo{pages}{012034}.
\bibitem[{Taylor(2008)}]{taylor2008non}
\bibinfo{author}{M.~Taylor},
\newblock \bibinfo{title}{Non-relativistic holography},
\newblock \bibinfo{journal}{arXiv preprint arXiv:0812.0530}
  (\bibinfo{year}{2008}).
\bibitem[{Griffin et~al.(2013)Griffin, Ho{\v{r}}ava, and
  Melby-Thompson}]{griffin2013lifshitz}
\bibinfo{author}{T.~Griffin}, \bibinfo{author}{P.~Ho{\v{r}}ava},
  \bibinfo{author}{C.~M. Melby-Thompson},
\newblock \bibinfo{title}{Lifshitz gravity for lifshitz holography},
\newblock \bibinfo{journal}{Physical review letters} \bibinfo{volume}{110}
  (\bibinfo{year}{2013}) \bibinfo{pages}{081602}.
\bibitem[{Balasubramanian and McGreevy(2009)}]{balasubramanian2009analytic}
\bibinfo{author}{K.~Balasubramanian}, \bibinfo{author}{J.~McGreevy},
\newblock \bibinfo{title}{An analytic lifshitz black hole},
\newblock \bibinfo{journal}{Physical Review D} \bibinfo{volume}{80}
  (\bibinfo{year}{2009}) \bibinfo{pages}{104039}.
\bibitem[{Danielsson and Thorlacius(2009)}]{danielsson2009black}
\bibinfo{author}{U.~H. Danielsson}, \bibinfo{author}{L.~Thorlacius},
\newblock \bibinfo{title}{Black holes in asymptotically lifshitz spacetime},
\newblock \bibinfo{journal}{Journal of High Energy Physics}
  \bibinfo{volume}{2009} (\bibinfo{year}{2009}) \bibinfo{pages}{070}.
\bibitem[{Mann(2009)}]{mann2009lifshitz}
\bibinfo{author}{R.~B. Mann},
\newblock \bibinfo{title}{Lifshitz topological black holes},
\newblock \bibinfo{journal}{Journal of High Energy Physics}
  \bibinfo{volume}{2009} (\bibinfo{year}{2009}) \bibinfo{pages}{075}.
\bibitem[{Devecio{\u{g}}lu(2014)}]{deveciouglu2014lifshitz}
\bibinfo{author}{D.~O. Devecio{\u{g}}lu},
\newblock \bibinfo{title}{Lifshitz black holes in einstein-yang-mills theory},
\newblock \bibinfo{journal}{Physical Review D} \bibinfo{volume}{89}
  (\bibinfo{year}{2014}) \bibinfo{pages}{124020}.
\bibitem[{Bertoldi et~al.(2009)Bertoldi, Burrington, and
  Peet}]{bertoldi2009black}
\bibinfo{author}{G.~Bertoldi}, \bibinfo{author}{B.~A. Burrington},
  \bibinfo{author}{A.~Peet},
\newblock \bibinfo{title}{Black holes in asymptotically lifshitz spacetimes
  with arbitrary critical exponent},
\newblock \bibinfo{journal}{Physical Review D} \bibinfo{volume}{80}
  (\bibinfo{year}{2009}) \bibinfo{pages}{126003}.
\bibitem[{Pang(2010)}]{pang2010charged}
\bibinfo{author}{D.-W. Pang},
\newblock \bibinfo{title}{On charged lifshitz black holes},
\newblock \bibinfo{journal}{Journal of High Energy Physics}
  \bibinfo{volume}{2010} (\bibinfo{year}{2010}) \bibinfo{pages}{1--20}.
\bibitem[{Brynjolfsson et~al.(2013)Brynjolfsson, Danielsson, Thorlacius, and
  Zingg}]{brynjolfsson2013holographic}
\bibinfo{author}{E.~Brynjolfsson}, \bibinfo{author}{U.~Danielsson},
  \bibinfo{author}{L.~Thorlacius}, \bibinfo{author}{T.~Zingg},
\newblock \bibinfo{title}{Holographic models with anisotropic scaling},
\newblock in: \bibinfo{booktitle}{Journal of Physics: Conference Series},
  volume \bibinfo{volume}{462}, \bibinfo{organization}{IOP Publishing},
  \bibinfo{year}{2013}, p. \bibinfo{pages}{012055}.
\bibitem[{Dehghani et~al.(2011)Dehghani, Mann, and
  Pourhasan}]{dehghani2011charged}
\bibinfo{author}{M.~H. Dehghani}, \bibinfo{author}{R.~B. Mann},
  \bibinfo{author}{R.~Pourhasan},
\newblock \bibinfo{title}{Charged lifshitz black holes},
\newblock \bibinfo{journal}{Physical Review D} \bibinfo{volume}{84}
  (\bibinfo{year}{2011}) \bibinfo{pages}{046002}.
\bibitem[{Alvarez et~al.(2014)Alvarez, Ay{\'o}n-Beato, Gonz{\'a}lez, and
  Hassa{\"\i}ne}]{alvarez2014nonlinearly}
\bibinfo{author}{A.~Alvarez}, \bibinfo{author}{E.~Ay{\'o}n-Beato},
  \bibinfo{author}{H.~A. Gonz{\'a}lez}, \bibinfo{author}{M.~Hassa{\"\i}ne},
\newblock \bibinfo{title}{Nonlinearly charged lifshitz black holes for any
  exponent z> 1},
\newblock \bibinfo{journal}{Journal of High Energy Physics}
  \bibinfo{volume}{2014} (\bibinfo{year}{2014}) \bibinfo{pages}{1--18}.
\bibitem[{Ay{\'o}n-Beato et~al.(2010)Ay{\'o}n-Beato, Garbarz, Giribet, and
  Hassaine}]{ayon2010analytic}
\bibinfo{author}{E.~Ay{\'o}n-Beato}, \bibinfo{author}{A.~Garbarz},
  \bibinfo{author}{G.~Giribet}, \bibinfo{author}{M.~Hassaine},
\newblock \bibinfo{title}{Analytic lifshitz black holes in higher dimensions},
\newblock \bibinfo{journal}{Journal of High Energy Physics}
  \bibinfo{volume}{2010} (\bibinfo{year}{2010}) \bibinfo{pages}{1--13}.
\bibitem[{Ay{\'o}n-Beato et~al.(2009)Ay{\'o}n-Beato, Garbarz, Giribet, and
  Hassaine}]{ayon2009lifshitz}
\bibinfo{author}{E.~Ay{\'o}n-Beato}, \bibinfo{author}{A.~Garbarz},
  \bibinfo{author}{G.~Giribet}, \bibinfo{author}{M.~Hassaine},
\newblock \bibinfo{title}{Lifshitz black hole in three dimensions},
\newblock \bibinfo{journal}{Physical Review D} \bibinfo{volume}{80}
  (\bibinfo{year}{2009}) \bibinfo{pages}{104029}.
\bibitem[{Brito and Santos(2020)}]{brito2020black}
\bibinfo{author}{F.~Brito}, \bibinfo{author}{F.~Santos},
\newblock \bibinfo{title}{Black brane in asymptotically lifshitz spacetime and
  viscosity/entropy ratios in horndeski gravity},
\newblock \bibinfo{journal}{EPL (Europhysics Letters)} \bibinfo{volume}{129}
  (\bibinfo{year}{2020}) \bibinfo{pages}{50003}.
\bibitem[{Natsuume and Okamura(2018)}]{natsuume2018holographic}
\bibinfo{author}{M.~Natsuume}, \bibinfo{author}{T.~Okamura},
\newblock \bibinfo{title}{Holographic lifshitz superconductors: analytic
  solution},
\newblock \bibinfo{journal}{Physical Review D} \bibinfo{volume}{97}
  (\bibinfo{year}{2018}) \bibinfo{pages}{066016}.
\bibitem[{Bazeia et~al.(2015)Bazeia, Brito, and Costa}]{bazeia2015two}
\bibinfo{author}{D.~Bazeia}, \bibinfo{author}{F.~A. Brito},
  \bibinfo{author}{F.~G. Costa},
\newblock \bibinfo{title}{Two-dimensional horava-lifshitz black hole
  solutions},
\newblock \bibinfo{journal}{Physical Review D} \bibinfo{volume}{91}
  (\bibinfo{year}{2015}) \bibinfo{pages}{044026}.
\bibitem[{Devecio{\u{g}}lu and
  Sar{\i}o{\u{g}}lu(2011)}]{deveciouglu2011thermodynamics}
\bibinfo{author}{D.~O. Devecio{\u{g}}lu},
  \bibinfo{author}{{\"O}.~Sar{\i}o{\u{g}}lu},
\newblock \bibinfo{title}{Thermodynamics of lifshitz black holes},
\newblock \bibinfo{journal}{Physical Review D} \bibinfo{volume}{83}
  (\bibinfo{year}{2011}) \bibinfo{pages}{124041}.
\bibitem[{Gonzalez et~al.(2011)Gonzalez, Tempo, and
  Troncoso}]{gonzalez2011field}
\bibinfo{author}{H.~A. Gonzalez}, \bibinfo{author}{D.~Tempo},
  \bibinfo{author}{R.~Troncoso},
\newblock \bibinfo{title}{Field theories with anisotropic scaling in 2d,
  solitons and the microscopic entropy of asymptotically lifshitz black holes},
\newblock \bibinfo{journal}{Journal of High Energy Physics}
  \bibinfo{volume}{2011} (\bibinfo{year}{2011}) \bibinfo{pages}{1--15}.
\bibitem[{Melnikov et~al.(2019)Melnikov, Novaes, P{\'e}rez, and
  Troncoso}]{melnikov2019lifshitz}
\bibinfo{author}{D.~Melnikov}, \bibinfo{author}{F.~Novaes},
  \bibinfo{author}{A.~P{\'e}rez}, \bibinfo{author}{R.~Troncoso},
\newblock \bibinfo{title}{Lifshitz scaling, microstate counting from number
  theory and black hole entropy},
\newblock \bibinfo{journal}{Journal of High Energy Physics}
  \bibinfo{volume}{2019} (\bibinfo{year}{2019}) \bibinfo{pages}{1--19}.
\bibitem[{Bravo-Gaete and Ju{\'a}rez-Aubry(2020)}]{bravo2020thermodynamics}
\bibinfo{author}{M.~Bravo-Gaete}, \bibinfo{author}{M.~M. Ju{\'a}rez-Aubry},
\newblock \bibinfo{title}{Thermodynamics and cardy-like formula for
  nonminimally dressed, charged lifshitz black holes in new massive gravity},
\newblock \bibinfo{journal}{Classical and Quantum Gravity} \bibinfo{volume}{37}
  (\bibinfo{year}{2020}) \bibinfo{pages}{075016}.
\bibitem[{Ay{\'o}n-Beato et~al.(2019)Ay{\'o}n-Beato, Bravo-Gaete, Correa,
  Hassaine, and Ju{\'a}rez-Aubry}]{ayon2019microscopic}
\bibinfo{author}{E.~Ay{\'o}n-Beato}, \bibinfo{author}{M.~Bravo-Gaete},
  \bibinfo{author}{F.~Correa}, \bibinfo{author}{M.~Hassaine},
  \bibinfo{author}{M.~M. Ju{\'a}rez-Aubry},
\newblock \bibinfo{title}{Microscopic entropy of higher-dimensional
  nonminimally dressed lifshitz black holes},
\newblock \bibinfo{journal}{Physical Review D} \bibinfo{volume}{100}
  (\bibinfo{year}{2019}) \bibinfo{pages}{044024}.
\bibitem[{Zangeneh et~al.(2015)Zangeneh, Sheykhi, and
  Dehghani}]{zangeneh2015thermodynamics}
\bibinfo{author}{M.~K. Zangeneh}, \bibinfo{author}{A.~Sheykhi},
  \bibinfo{author}{M.~H. Dehghani},
\newblock \bibinfo{title}{Thermodynamics of topological nonlinear charged
  lifshitz black holes},
\newblock \bibinfo{journal}{Physical Review D} \bibinfo{volume}{92}
  (\bibinfo{year}{2015}) \bibinfo{pages}{024050}.
\bibitem[{Liu and L{\"u}(2014)}]{liu2014thermodynamics}
\bibinfo{author}{H.-S. Liu}, \bibinfo{author}{H.~L{\"u}},
\newblock \bibinfo{title}{Thermodynamics of lifshitz black holes},
\newblock \bibinfo{journal}{Journal of High Energy Physics}
  \bibinfo{volume}{2014} (\bibinfo{year}{2014}) \bibinfo{pages}{1--24}.
\bibitem[{Tarrio and Vandoren(2011)}]{tarrio2011black}
\bibinfo{author}{J.~Tarrio}, \bibinfo{author}{S.~Vandoren},
\newblock \bibinfo{title}{Black holes and black branes in lifshitz spacetimes},
\newblock \bibinfo{journal}{Journal of High Energy Physics}
  \bibinfo{volume}{2011} (\bibinfo{year}{2011}) \bibinfo{pages}{1--27}.
\bibitem[{Naeimipour et~al.(2021)Naeimipour, Mirza, and
  Nasirimoghadam}]{naeimipour2021lifshitz}
\bibinfo{author}{F.~Naeimipour}, \bibinfo{author}{B.~Mirza},
  \bibinfo{author}{S.~Nasirimoghadam},
\newblock \bibinfo{title}{Lifshitz and hyperscaling violated yang-mills-dilaton
  black holes},
\newblock \bibinfo{journal}{Physical Review D} \bibinfo{volume}{104}
  (\bibinfo{year}{2021}) \bibinfo{pages}{104059}.
\bibitem[{Tallarita(2014)}]{tallarita2014holographic}
\bibinfo{author}{G.~Tallarita},
\newblock \bibinfo{title}{Holographic lifshitz superconductors with an axion
  field},
\newblock \bibinfo{journal}{Physical Review D} \bibinfo{volume}{89}
  (\bibinfo{year}{2014}) \bibinfo{pages}{106005}.
\bibitem[{Li et~al.(2014)Li, Li, Wang, and Zhang}]{li2014non}
\bibinfo{author}{H.-F. Li}, \bibinfo{author}{L.~Li}, \bibinfo{author}{Y.-Q.
  Wang}, \bibinfo{author}{H.-Q. Zhang},
\newblock \bibinfo{title}{Non-relativistic josephson junction from holography},
\newblock \bibinfo{journal}{Journal of High Energy Physics}
  \bibinfo{volume}{2014} (\bibinfo{year}{2014}) \bibinfo{pages}{1--15}.
\bibitem[{Lu et~al.(2014)Lu, Wu, Qian, Zhao, Zhang, and Zhang}]{lu2014lifshitz}
\bibinfo{author}{J.-W. Lu}, \bibinfo{author}{Y.-B. Wu},
  \bibinfo{author}{P.~Qian}, \bibinfo{author}{Y.-Y. Zhao},
  \bibinfo{author}{X.~Zhang}, \bibinfo{author}{N.~Zhang},
\newblock \bibinfo{title}{Lifshitz scaling effects on holographic
  superconductors},
\newblock \bibinfo{journal}{Nuclear Physics B} \bibinfo{volume}{887}
  (\bibinfo{year}{2014}) \bibinfo{pages}{112--135}.
\bibitem[{Bravo-Gaete et~al.(2022)Bravo-Gaete, Ju{\'a}rez-Aubry, and
  Rodr{\'\i}guez}]{bravo2022lifshitz}
\bibinfo{author}{M.~Bravo-Gaete}, \bibinfo{author}{M.~M. Ju{\'a}rez-Aubry},
  \bibinfo{author}{G.~V. Rodr{\'\i}guez},
\newblock \bibinfo{title}{Lifshitz black holes in four-dimensional critical
  gravity},
\newblock \bibinfo{journal}{Physical Review D} \bibinfo{volume}{105}
  (\bibinfo{year}{2022}) \bibinfo{pages}{084009}.
\bibitem[{Chamblin et~al.(1999{\natexlab{a}})Chamblin, Emparan, Johnson, and
  Myers}]{chamblin1999charged}
\bibinfo{author}{A.~Chamblin}, \bibinfo{author}{R.~Emparan},
  \bibinfo{author}{C.~V. Johnson}, \bibinfo{author}{R.~C. Myers},
\newblock \bibinfo{title}{Charged ads black holes and catastrophic holography},
\newblock \bibinfo{journal}{Physical Review D} \bibinfo{volume}{60}
  (\bibinfo{year}{1999}{\natexlab{a}}) \bibinfo{pages}{064018}.
\bibitem[{Chamblin et~al.(1999{\natexlab{b}})Chamblin, Emparan, Johnson, and
  Myers}]{chamblin1999holography}
\bibinfo{author}{A.~Chamblin}, \bibinfo{author}{R.~Emparan},
  \bibinfo{author}{C.~V. Johnson}, \bibinfo{author}{R.~C. Myers},
\newblock \bibinfo{title}{Holography, thermodynamics, and fluctuations of
  charged ads black holes},
\newblock \bibinfo{journal}{Physical Review D} \bibinfo{volume}{60}
  (\bibinfo{year}{1999}{\natexlab{b}}) \bibinfo{pages}{104026}.
\bibitem[{Dolan(2011)}]{dolan2011pressure}
\bibinfo{author}{B.~P. Dolan},
\newblock \bibinfo{title}{Pressure and volume in the first law of black hole
  thermodynamics},
\newblock \bibinfo{journal}{Classical and Quantum Gravity} \bibinfo{volume}{28}
  (\bibinfo{year}{2011}) \bibinfo{pages}{235017}.
\bibitem[{Kubiz{\v{n}}{\'a}k and Mann(2012)}]{kubizvnak2012p}
\bibinfo{author}{D.~Kubiz{\v{n}}{\'a}k}, \bibinfo{author}{R.~B. Mann},
\newblock \bibinfo{title}{P- v criticality of charged ads black holes},
\newblock \bibinfo{journal}{Journal of High Energy Physics}
  \bibinfo{volume}{2012} (\bibinfo{year}{2012}) \bibinfo{pages}{1--25}.
\bibitem[{Brynj{\'o}lfsson et~al.(2010)Brynj{\'o}lfsson, Danielsson,
  Thorlacius, and Zingg}]{brynjolfsson2010holographic}
\bibinfo{author}{E.~J. Brynj{\'o}lfsson}, \bibinfo{author}{U.~H. Danielsson},
  \bibinfo{author}{L.~Thorlacius}, \bibinfo{author}{T.~Zingg},
\newblock \bibinfo{title}{Holographic superconductors with lifshitz scaling},
\newblock \bibinfo{journal}{Journal of Physics A: Mathematical and Theoretical}
  \bibinfo{volume}{43} (\bibinfo{year}{2010}) \bibinfo{pages}{065401}.
\bibitem[{Zhu et~al.(2020)Zhu, Shu, and Du}]{zhu2020holographic}
\bibinfo{author}{K.-X. Zhu}, \bibinfo{author}{F.-W. Shu},
  \bibinfo{author}{D.-H. Du},
\newblock \bibinfo{title}{Holographic complexity for nonlinearly charged
  lifshitz black holes},
\newblock \bibinfo{journal}{Classical and Quantum Gravity} \bibinfo{volume}{37}
  (\bibinfo{year}{2020}) \bibinfo{pages}{195023}.
\bibitem[{Jain(2010)}]{jain2010universal}
\bibinfo{author}{S.~Jain},
\newblock \bibinfo{title}{Universal thermal and electrical conductivity from
  holography},
\newblock \bibinfo{journal}{Journal of High Energy Physics}
  \bibinfo{volume}{2010} (\bibinfo{year}{2010}) \bibinfo{pages}{1--32}.
\bibitem[{Brynjolfsson et~al.(2010)Brynjolfsson, Danielsson, Thorlaciusa, and
  Zingg}]{brynjolfsson2010black}
\bibinfo{author}{E.~Brynjolfsson}, \bibinfo{author}{U.~H. Danielsson},
  \bibinfo{author}{L.~Thorlaciusa}, \bibinfo{author}{T.~Zingg},
\newblock \bibinfo{title}{Black hole thermodynamics and heavy fermion metals},
\newblock \bibinfo{journal}{Journal of High Energy Physics}
  \bibinfo{volume}{2010} (\bibinfo{year}{2010}) \bibinfo{pages}{1--16}.
\bibitem[{Fang et~al.(2012)Fang, Ge, and Kuang}]{fang2012holographic}
\bibinfo{author}{L.~Q. Fang}, \bibinfo{author}{X.-H. Ge},
  \bibinfo{author}{X.-M. Kuang},
\newblock \bibinfo{title}{Holographic fermions in charged lifshitz theory},
\newblock \bibinfo{journal}{Physical Review D} \bibinfo{volume}{86}
  (\bibinfo{year}{2012}) \bibinfo{pages}{105037}.
\bibitem[{Hobart(1963)}]{hobart1963instability}
\bibinfo{author}{R.~Hobart},
\newblock \bibinfo{title}{On the instability of a class of unitary field
  models},
\newblock \bibinfo{journal}{Proceedings of the Physical Society (1958-1967)}
  \bibinfo{volume}{82} (\bibinfo{year}{1963}) \bibinfo{pages}{201}.
\bibitem[{Derrick(1964)}]{derrick1964comments}
\bibinfo{author}{G.~Derrick},
\newblock \bibinfo{title}{Comments on nonlinear wave equations as models for
  elementary particles},
\newblock \bibinfo{journal}{Journal of Mathematical Physics}
  \bibinfo{volume}{5} (\bibinfo{year}{1964}) \bibinfo{pages}{1252--1254}.
\bibitem[{Radmore and Stephenson(1978)}]{radmore1978non}
\bibinfo{author}{P.~Radmore}, \bibinfo{author}{G.~Stephenson},
\newblock \bibinfo{title}{Non-linear wave equations in a curved background
  space},
\newblock \bibinfo{journal}{Journal of Physics A: Mathematical and General}
  \bibinfo{volume}{11} (\bibinfo{year}{1978}) \bibinfo{pages}{L149}.
\bibitem[{Palmer(1979)}]{palmer1979derrick}
\bibinfo{author}{T.~Palmer},
\newblock \bibinfo{title}{Derrick's theorem in curved space},
\newblock \bibinfo{journal}{Journal of Physics A: Mathematical and General}
  \bibinfo{volume}{12} (\bibinfo{year}{1979}) \bibinfo{pages}{L17}.
\bibitem[{Bazeia et~al.(2003)Bazeia, Menezes, and Menezes}]{bazeia2003new}
\bibinfo{author}{D.~Bazeia}, \bibinfo{author}{J.~Menezes},
  \bibinfo{author}{R.~Menezes},
\newblock \bibinfo{title}{New global defect structures},
\newblock \bibinfo{journal}{Physical Review Letters} \bibinfo{volume}{91}
  (\bibinfo{year}{2003}) \bibinfo{pages}{241601}.
\bibitem[{Alestas and Perivolaropoulos(2019)}]{alestas2019evading}
\bibinfo{author}{G.~Alestas}, \bibinfo{author}{L.~Perivolaropoulos},
\newblock \bibinfo{title}{Evading derrick’s theorem in curved space: Static
  metastable spherical domain wall},
\newblock \bibinfo{journal}{Physical Review D} \bibinfo{volume}{99}
  (\bibinfo{year}{2019}) \bibinfo{pages}{064026}.
\bibitem[{Carloni and Rosa(2019)}]{carloni2019derrick}
\bibinfo{author}{S.~Carloni}, \bibinfo{author}{J.~L. Rosa},
\newblock \bibinfo{title}{Derrick’s theorem in curved spacetime},
\newblock \bibinfo{journal}{Physical Review D} \bibinfo{volume}{100}
  (\bibinfo{year}{2019}) \bibinfo{pages}{025014}.
\bibitem[{Morris(2021)}]{morris2021radially}
\bibinfo{author}{J.~R. Morris},
\newblock \bibinfo{title}{Radially symmetric scalar solitons},
\newblock \bibinfo{journal}{Physical Review D} \bibinfo{volume}{104}
  (\bibinfo{year}{2021}) \bibinfo{pages}{016013}.
\bibitem[{Mandal(2021)}]{mandal2021solitons}
\bibinfo{author}{S.~Mandal},
\newblock \bibinfo{title}{Solitons in curved spacetime},
\newblock \bibinfo{journal}{Europhysics Letters} \bibinfo{volume}{136}
  (\bibinfo{year}{2021}) \bibinfo{pages}{11001}.
\bibitem[{Morris(2022)}]{morris2022bps}
\bibinfo{author}{J.~Morris},
\newblock \bibinfo{title}{Bps equations and solutions for maxwell--scalar
  theory},
\newblock \bibinfo{journal}{Annals of Physics} \bibinfo{volume}{438}
  (\bibinfo{year}{2022}) \bibinfo{pages}{168782}.
\bibitem[{Moreira(2022{\natexlab{a}})}]{moreira2022analytical}
\bibinfo{author}{D.~C. Moreira},
\newblock \bibinfo{title}{Analytical scalar field solutions on lifshitz
  spacetimes},
\newblock \bibinfo{journal}{Physical Review D} \bibinfo{volume}{105}
  (\bibinfo{year}{2022}{\natexlab{a}}) \bibinfo{pages}{016001}.
\bibitem[{Moreira(2022{\natexlab{b}})}]{moreira2022erratum}
\bibinfo{author}{D.~C. Moreira},
\newblock \bibinfo{title}{Erratum: Analytical scalar field solutions on
  lifshitz spacetimes [phys. rev. d 105, 016001 (2022)]},
\newblock \bibinfo{journal}{Physical Review D} \bibinfo{volume}{106}
  (\bibinfo{year}{2022}{\natexlab{b}}) \bibinfo{pages}{039903(E)}.
\bibitem[{Moreira et~al.(2022)Moreira, Brito, and
  Mota-Silva}]{moreira2022scalar}
\bibinfo{author}{D.~C. Moreira}, \bibinfo{author}{F.~A. Brito},
  \bibinfo{author}{J.~C. Mota-Silva},
\newblock \bibinfo{title}{Scalar fields and lifshitz black holes from
  derrick’s theorem evasion},
\newblock \bibinfo{journal}{Physical Review D} \bibinfo{volume}{106}
  (\bibinfo{year}{2022}) \bibinfo{pages}{125017}.
\bibitem[{Moreira et~al.(2023)Moreira, Brito, and
  Bazeia}]{moreira2023localized}
\bibinfo{author}{D.~Moreira}, \bibinfo{author}{F.~Brito},
  \bibinfo{author}{D.~Bazeia},
\newblock \bibinfo{title}{Localized scalar structures around static black
  holes},
\newblock \bibinfo{journal}{Nuclear Physics B} \bibinfo{volume}{987}
  (\bibinfo{year}{2023}) \bibinfo{pages}{116090}.
\bibitem[{Anber et~al.(2010)Anber, Aydemir, and Donoghue}]{anber2010breaking}
\bibinfo{author}{M.~M. Anber}, \bibinfo{author}{U.~Aydemir},
  \bibinfo{author}{J.~F. Donoghue},
\newblock \bibinfo{title}{Breaking diffeomorphism invariance and tests for the
  emergence of gravity},
\newblock \bibinfo{journal}{Physical Review D} \bibinfo{volume}{81}
  (\bibinfo{year}{2010}) \bibinfo{pages}{084059}.
\bibitem[{Cannone et~al.(2015)Cannone, Tasinato, and
  Wands}]{cannone2015generalised}
\bibinfo{author}{D.~Cannone}, \bibinfo{author}{G.~Tasinato},
  \bibinfo{author}{D.~Wands},
\newblock \bibinfo{title}{Generalised tensor fluctuations and inflation},
\newblock \bibinfo{journal}{Journal of Cosmology and Astroparticle Physics}
  \bibinfo{volume}{2015} (\bibinfo{year}{2015}) \bibinfo{pages}{029}.
\bibitem[{Graef and Brandenberger(2015)}]{graef2015breaking}
\bibinfo{author}{L.~Graef}, \bibinfo{author}{R.~Brandenberger},
\newblock \bibinfo{title}{Breaking of spatial diffeomorphism invariance,
  inflation and the spectrum of cosmological perturbations},
\newblock \bibinfo{journal}{Journal of Cosmology and Astroparticle Physics}
  \bibinfo{volume}{2015} (\bibinfo{year}{2015}) \bibinfo{pages}{009}.
\bibitem[{Graef et~al.(2017)Graef, Benetti, and
  Alcaniz}]{graef2017constraining}
\bibinfo{author}{L.~Graef}, \bibinfo{author}{M.~Benetti},
  \bibinfo{author}{J.~Alcaniz},
\newblock \bibinfo{title}{Constraining the break of spatial diffeomorphism
  invariance with planck data},
\newblock \bibinfo{journal}{Journal of Cosmology and Astroparticle Physics}
  \bibinfo{volume}{2017} (\bibinfo{year}{2017}) \bibinfo{pages}{013}.
\bibitem[{Milgrom(2019)}]{milgrom2019noncovariance}
\bibinfo{author}{M.~Milgrom},
\newblock \bibinfo{title}{Noncovariance at low accelerations as a route to
  mond},
\newblock \bibinfo{journal}{Physical Review D} \bibinfo{volume}{100}
  (\bibinfo{year}{2019}) \bibinfo{pages}{084039}.
\bibitem[{Reyes and Schreck(2021)}]{reyes2021hamiltonian}
\bibinfo{author}{C.~M. Reyes}, \bibinfo{author}{M.~Schreck},
\newblock \bibinfo{title}{Hamiltonian formulation of an effective modified
  gravity with nondynamical background fields},
\newblock \bibinfo{journal}{Physical Review D} \bibinfo{volume}{104}
  (\bibinfo{year}{2021}) \bibinfo{pages}{124042}.
\bibitem[{Reyes and Schreck(2022)}]{reyes2022modified}
\bibinfo{author}{C.~M. Reyes}, \bibinfo{author}{M.~Schreck},
\newblock \bibinfo{title}{Modified-gravity theories with nondynamical
  background fields},
\newblock \bibinfo{journal}{Physical Review D} \bibinfo{volume}{106}
  (\bibinfo{year}{2022}) \bibinfo{pages}{044050}.
\bibitem[{Reyes et~al.(2022)Reyes, Schreck, and Soto}]{reyes2022cosmology}
\bibinfo{author}{C.~M. Reyes}, \bibinfo{author}{M.~Schreck},
  \bibinfo{author}{A.~Soto},
\newblock \bibinfo{title}{Cosmology in the presence of
  diffeomorphism-violating, nondynamical background fields},
\newblock \bibinfo{journal}{Physical Review D} \bibinfo{volume}{106}
  (\bibinfo{year}{2022}) \bibinfo{pages}{023524}.
\bibitem[{Kosteleck{\`y} and Li(2021)}]{kostelecky2021backgrounds}
\bibinfo{author}{V.~A. Kosteleck{\`y}}, \bibinfo{author}{Z.~Li},
\newblock \bibinfo{title}{Backgrounds in gravitational effective field theory},
\newblock \bibinfo{journal}{Physical Review D} \bibinfo{volume}{103}
  (\bibinfo{year}{2021}) \bibinfo{pages}{024059}.
\bibitem[{Bluhm(2017)}]{bluhm2017gravity}
\bibinfo{author}{R.~Bluhm},
\newblock \bibinfo{title}{Gravity theories with background fields and spacetime
  symmetry breaking},
\newblock \bibinfo{journal}{Symmetry} \bibinfo{volume}{9}
  (\bibinfo{year}{2017}) \bibinfo{pages}{230}.
\bibitem[{Bluhm(2015)}]{bluhm2015spacetime}
\bibinfo{author}{R.~Bluhm},
\newblock \bibinfo{title}{Spacetime symmetry breaking and einstein-maxwell
  theory},
\newblock \bibinfo{journal}{Physical Review D} \bibinfo{volume}{92}
  (\bibinfo{year}{2015}) \bibinfo{pages}{085015}.
\bibitem[{Bluhm et~al.(2019)Bluhm, Bossi, and Wen}]{bluhm2019gravity}
\bibinfo{author}{R.~Bluhm}, \bibinfo{author}{H.~Bossi},
  \bibinfo{author}{Y.~Wen},
\newblock \bibinfo{title}{Gravity with explicit spacetime symmetry breaking and
  the standard model extension},
\newblock \bibinfo{journal}{Physical Review D} \bibinfo{volume}{100}
  (\bibinfo{year}{2019}) \bibinfo{pages}{084022}.
\bibitem[{Bluhm and Yang(2021)}]{bluhm2021gravity}
\bibinfo{author}{R.~Bluhm}, \bibinfo{author}{Y.~Yang},
\newblock \bibinfo{title}{Gravity with explicit diffeomorphism breaking},
\newblock \bibinfo{journal}{Symmetry} \bibinfo{volume}{13}
  (\bibinfo{year}{2021}) \bibinfo{pages}{660}.
\bibitem[{Hidaka et~al.(2015)Hidaka, Noumi, and Shiu}]{hidaka2015effective}
\bibinfo{author}{Y.~Hidaka}, \bibinfo{author}{T.~Noumi},
  \bibinfo{author}{G.~Shiu},
\newblock \bibinfo{title}{Effective field theory for spacetime symmetry
  breaking},
\newblock \bibinfo{journal}{Physical Review D} \bibinfo{volume}{92}
  (\bibinfo{year}{2015}) \bibinfo{pages}{045020}.
\bibitem[{Bluhm(2015)}]{bluhm2015explicit}
\bibinfo{author}{R.~Bluhm},
\newblock \bibinfo{title}{Explicit versus spontaneous diffeomorphism breaking
  in gravity},
\newblock \bibinfo{journal}{Physical Review D} \bibinfo{volume}{91}
  (\bibinfo{year}{2015}) \bibinfo{pages}{065034}.
\bibitem[{Kastor et~al.(2009)Kastor, Ray, and Traschen}]{kastor2009enthalpy}
\bibinfo{author}{D.~Kastor}, \bibinfo{author}{S.~Ray},
  \bibinfo{author}{J.~Traschen},
\newblock \bibinfo{title}{Enthalpy and the mechanics of ads black holes},
\newblock \bibinfo{journal}{Classical and Quantum Gravity} \bibinfo{volume}{26}
  (\bibinfo{year}{2009}) \bibinfo{pages}{195011}.
\bibitem[{Dolan(2011)}]{dolan2011cosmological}
\bibinfo{author}{B.~P. Dolan},
\newblock \bibinfo{title}{The cosmological constant and black-hole
  thermodynamic potentials},
\newblock \bibinfo{journal}{Classical and Quantum Gravity} \bibinfo{volume}{28}
  (\bibinfo{year}{2011}) \bibinfo{pages}{125020}.
\bibitem[{Kubiz{\v{n}}{\'a}k et~al.(2017)Kubiz{\v{n}}{\'a}k, Mann, and
  Teo}]{kubizvnak2017black}
\bibinfo{author}{D.~Kubiz{\v{n}}{\'a}k}, \bibinfo{author}{R.~B. Mann},
  \bibinfo{author}{M.~Teo},
\newblock \bibinfo{title}{Black hole chemistry: thermodynamics with lambda},
\newblock \bibinfo{journal}{Classical and Quantum Gravity} \bibinfo{volume}{34}
  (\bibinfo{year}{2017}) \bibinfo{pages}{063001}.
\bibitem[{Hartnoll et~al.(2020)Hartnoll, Horowitz, Kruthoff, and
  Santos}]{hartnoll2020gravitational}
\bibinfo{author}{S.~A. Hartnoll}, \bibinfo{author}{G.~T. Horowitz},
  \bibinfo{author}{J.~Kruthoff}, \bibinfo{author}{J.~E. Santos},
\newblock \bibinfo{title}{Gravitational duals to the grand canonical ensemble
  abhor cauchy horizons},
\newblock \bibinfo{journal}{Journal of High Energy Physics}
  \bibinfo{volume}{2020} (\bibinfo{year}{2020}) \bibinfo{pages}{1--24}.
\bibitem[{Hartnoll et~al.(2021)Hartnoll, Horowitz, Kruthoff, and
  Santos}]{hartnoll2021diving}
\bibinfo{author}{S.~A. Hartnoll}, \bibinfo{author}{G.~Horowitz},
  \bibinfo{author}{J.~Kruthoff}, \bibinfo{author}{J.~Santos},
\newblock \bibinfo{title}{Diving into a holographic superconductor},
\newblock \bibinfo{journal}{SciPost Physics} \bibinfo{volume}{10}
  (\bibinfo{year}{2021}) \bibinfo{pages}{009}.
\bibitem[{Cai et~al.(2021)Cai, Li, and Yang}]{cai2021no}
\bibinfo{author}{R.-G. Cai}, \bibinfo{author}{L.~Li}, \bibinfo{author}{R.-Q.
  Yang},
\newblock \bibinfo{title}{No inner-horizon theorem for black holes with charged
  scalar hairs},
\newblock \bibinfo{journal}{Journal of High Energy Physics}
  \bibinfo{volume}{2021} (\bibinfo{year}{2021}) \bibinfo{pages}{1--26}.
\bibitem[{An et~al.(2021)An, Li, and Yang}]{an2021no}
\bibinfo{author}{Y.-S. An}, \bibinfo{author}{L.~Li}, \bibinfo{author}{F.-G.
  Yang},
\newblock \bibinfo{title}{No cauchy horizon theorem for nonlinear
  electrodynamics black holes with charged scalar hairs},
\newblock \bibinfo{journal}{Physical Review D} \bibinfo{volume}{104}
  (\bibinfo{year}{2021}) \bibinfo{pages}{024040}.
\bibitem[{Lemos(1995)}]{lemos1995three}
\bibinfo{author}{J.~S. Lemos},
\newblock \bibinfo{title}{Three dimensional black holes and cylindrical general
  relativity},
\newblock \bibinfo{journal}{Physics Letters B} \bibinfo{volume}{353}
  (\bibinfo{year}{1995}) \bibinfo{pages}{46--51}.
\bibitem[{D{\'a}rlla et~al.(2023)D{\'a}rlla, Brito, and
  Furtado}]{darlla2023black}
\bibinfo{author}{R.~D{\'a}rlla}, \bibinfo{author}{F.~A. Brito},
  \bibinfo{author}{J.~Furtado},
\newblock \bibinfo{title}{Black string solutions in rainbow gravity},
\newblock \bibinfo{journal}{Universe} \bibinfo{volume}{9}
  (\bibinfo{year}{2023}) \bibinfo{pages}{297}.
\bibitem[{Gunasekaran et~al.(2012)Gunasekaran, Kubiz{\v{n}}{\'a}k, and
  Mann}]{gunasekaran2012extended}
\bibinfo{author}{S.~Gunasekaran}, \bibinfo{author}{D.~Kubiz{\v{n}}{\'a}k},
  \bibinfo{author}{R.~B. Mann},
\newblock \bibinfo{title}{Extended phase space thermodynamics for charged and
  rotating black holes and born-infeld vacuum polarization},
\newblock \bibinfo{journal}{Journal of High Energy Physics}
  \bibinfo{volume}{2012} (\bibinfo{year}{2012}) \bibinfo{pages}{1--43}.
\bibitem[{Vargaftik(1975)}]{vargaftik1975tables}
\bibinfo{author}{N.~Vargaftik}, \bibinfo{title}{Tables on the Thermophysical
  Properties of Liquids and Gases: In Normal and Dissociated States}, A Halsted
  Press book, \bibinfo{publisher}{Hemisphere Publishing Corporation},
  \bibinfo{year}{1975}.

\end{thebibliography}
\end{document}